\documentclass[12pt]{article}
\usepackage{latexsym}
\usepackage{float}
\usepackage{amsmath,bm,amssymb}
\usepackage{colordvi}
\usepackage{multicol,multirow}
\usepackage{color}
\usepackage{rotating}
 \usepackage{url,hyperref}
\usepackage[ruled]{algorithm2e}
\usepackage{subfigure}
\usepackage{placeins}
\usepackage{tabularx}
\usepackage{listings}

\usepackage{breqn}
\usepackage{amsfonts}
\usepackage{hyperref}
\usepackage{longtable}
\usepackage[font={small}]{caption}

\def\doublespace{\baselineskip=22pt}
\usepackage{lscape}

\usepackage[super,sort&compress]{natbib}

\begin{document}
\doublespace
\baselineskip 2.8ex
\begin{center}
{\bf\Huge The spike-and-slab quantile LASSO for robust variable selection in cancer genomics studies }

\

{\bf Yuwen Liu$^{1}$, Jie Ren$^{2}$, Shuangge Ma$^{3}$ and Cen Wu$^{1}$}\\

{ $^1$ Department of Statistics, Kansas State University, Manhattan, KS}\\

{ $^2$ Department of Biostatistics and Health Data Sciences, Indiana University School of Medicine, Indianapolis, IN}\\

{ $^3$ Department of Biostatistics, Yale University, New Haven, CT }\\

\end{center}

{\bf $\ast$ Corresponding author}:
Cen Wu, wucen@ksu.edu\\
\vspace{0.8em}

\noindent {\bf\Large Abstract}\\

\noindent Data irregularity in cancer genomics studies has been widely observed in the form of outliers and heavy-tailed distributions in the complex traits. In the past decade, robust variable selection methods have emerged as powerful alternatives to the non-robust ones to identify important genes associated with heterogeneous disease traits and build superior predictive models. In this study, to keep the remarkable features of the quantile LASSO and fully Bayesian regularized quantile regression while overcoming their disadvantage in the analysis of high-dimensional genomics data, we propose the spike-and-slab quantile LASSO through a fully Bayesian spike-and-slab formulation under the robust likelihood by adopting the asymmetric Laplace distribution (ALD). The proposed robust method has inherited the prominent properties of selective shrinkage and self-adaptivity to the sparsity pattern from the spike-and-slab LASSO (Ro\u{c}kov\'a and George, 2018)\cite{rovckova2018spike}. Furthermore, the spike-and-slab quantile LASSO has a computational advantage to locate the posterior modes via soft-thresholding rule guided Expectation-Maximization (EM) steps in the coordinate descent framework, a phenomenon rarely observed for robust regularization with non-differentiable loss functions. We have conducted comprehensive simulation studies with a variety of heavy-tailed errors in both homogeneous and heterogeneous model settings to demonstrate the superiority of the spike-and-slab quantile LASSO over its competing methods. The advantage of the proposed method has been further demonstrated in case studies of the lung adenocarcinomas (LUAD) and skin cutaneous melanoma (SKCM) data from The Cancer Genome Atlas (TCGA).


\newpage
\section{Introduction}

\begin{figure}[h]
	\includegraphics[height=0.35\textheight]{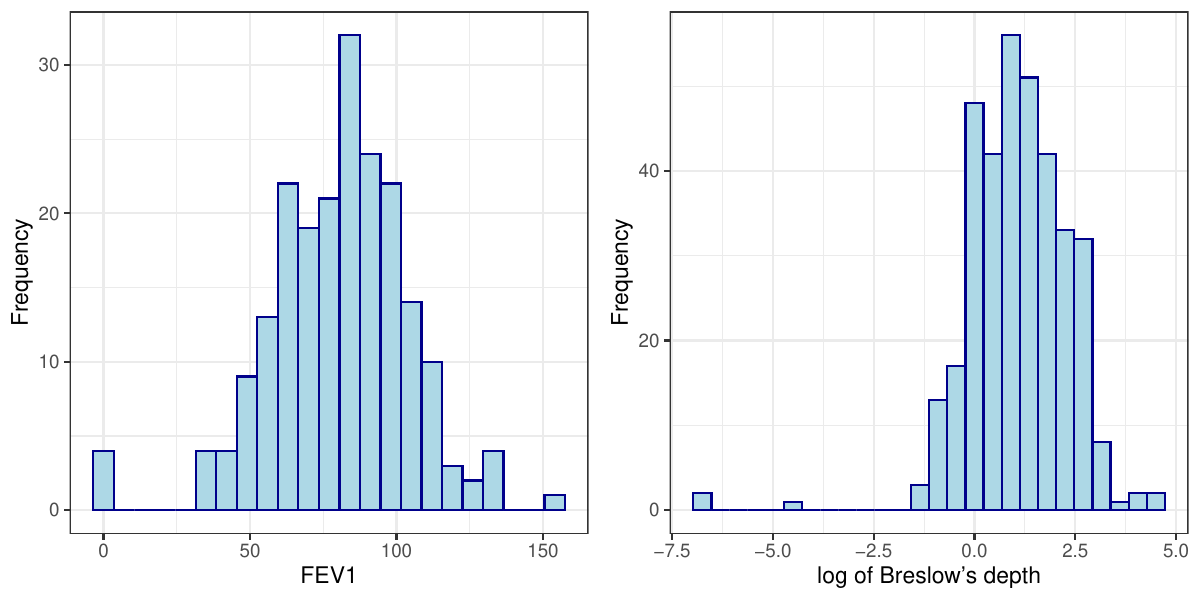}
	\caption{Histograms of FEV1 and log Breslow depth.}\label{hists}
\end{figure}

Modern high-throughput technologies have generated large scale omics data, including gene expressions, copy number variations, singlenucleotide polymorphisms (SNPs), that can be analyzed in cancer research to dissect the genetic and genomic architecture under the complex disease phenotypes. In practice, the heavy-tailed distributions and outliers have been widely observed in those disease traits, reflecting the heterogeneity of cancer. We take the examples from the The Cancer Genomics Atlas (TCGA) to illustrate the irregularities in the cancer phenotypes (Cerami et al., 2012)\cite{cerami2012cbio}. In the TCGA lung adenocarcinomas (LUAD) data (Campbell et al., 2011)\cite{campbell2016distinct}, the phenotype of interest is FEV1, the total volume of air expelled out of the lung in one second. Besides, in TCGA skin cutaneous melanoma (SKCM) study, we are interested in the log of Breslow’s thickness (Marghoob et al., 2000)\cite{marghoob2000breslow}, an indicator of the depth that melanoma tumor has grown into skin, as the disease trait. The histograms of the two phenotypes in Figure \ref{hists} have clearly indicated that they do not follow normal distribution, and are possibly contaminated by the outliers. The deviation from normality can be further confirmed by the p-values from  Shapiro-Wilk normality test, which are 0.003384 and 3.562$\times 10^{-14}$ for FEV1 and log Breslow depth, respectively. 

Identifying important disease-gene associations while building powerful predictive models under the disease phenotypes is the central task of analyzing high-throughput cancer genomics data. However, with non-robust methods, even a single outlying observation from the phenotypic traits will result in biased estimation and false detection of associated omics features \cite{wu2015selective}. Our exploratory data analysis has demonstrated that robust sparse regression is in pressing need for performing such a challenging task. In published studies, robust variable selection methods have been extensively developed in bioinformatics studies\cite{wu2015selective}, for the sparse identification of gene expressions \cite{alfons2013sparse,ren2019robust}, single nucleotide polymorphisms \cite{bradic2011penalized,daye2012high}\cite{peng2015iterative}, DNA copy number variations \cite{gao2010robust}\cite{wang2011identification,wu2018robust}, and proteins \cite{cohen2019robust,kepplinger2021robust}.

Within the Bayesian framework, robust variable selection has recently been developed in addition to widely examined nonrobust counterpart \cite{o2009review,ickstadt2018toward,johndrow2020scalable,liang2022adaptive,liang2023adaptive}. For example, multiple robust Bayesian methods have been proposed to identify important gene--environment (G$\times$E) interactions \cite{lu2021identifying,ren2023robust,zhou2023bayesian} and  histopathological imaging--environment interactions \cite{im2023bayesian}. In the Bayesian hierarchical model, incorporating the robustness against outliers and performing variable selection can be facilitated by adopting robust likelihood function and sparsity-inducing priors, respectively. For example, using the asymmetric Laplace distribution for the robust likelihood function \cite{yu2001bayesian}, Li et al. (2010)\cite{li2010bayesian} have established a Bayesian framework for regularization in linear quantile regression using the Laplace shrinkage priors. Other widely used sparse priors include the spike-and-slab priors \cite{george1993variable} and global local shrinkage priors such as the Normal Exponential Gamma prior \cite{brown2010inference} and Horseshoe prior \cite{carvalho2009handling,johndrow2020scalable}. Our biased literature search suggests that although there has been a surge of sparse Bayesian methods for genomics studies in the past decades \cite{o2009review,ickstadt2018toward}, the robustness in Bayesian analysis of omics data has not received as much attention as in those frequentist studies.

As the frequentist and Bayesian methods have their own pros and cons, the Spike-and-Slab LASSO (ssLASSO) has been proposed to retain remarkable properties of both while overcoming their respective limitations (Ro\u{c}kov\'a and George, 2018)\cite{rovckova2018spike}. Arising from a fully Bayes formulation with the spike-and-slab prior, ssLASSO leads to a posterior mode estimation algorithm via incorporating the Expectation-Maximization (EM) steps in the coordinate-descent framework \cite{friedman2010regularization}, and is thus extremely fast. The spike-and-slab LASSO has been furthered extended to handle survival outcomes with Cox models \cite{tang2017spike} and generalized responses including multinomial and binary disease outcomes \cite{tang2017spikeGLM}, demonstrating great promise in cancer genomics studies.    

Despite the success, a major limitation of spike-and-slab LASSO based methods is that they are non-robust and vulnerable to the heterogeneity of cancer, usually in the form of heavy-tailed distributions and outliers in the disease traits. On the other hand, robust variable selection methods face the computational hurdles due to the non-differentiable objective functions (including quantile check loss) for the frequentist ones \cite{wu2015selective} and computationally intensive MCMC algorithms for the fully Bayesian analysis. These limitations have motivated us to develop spike-and-slab quantile LASSO (ssQLASSO), which has a fully Bayes spike-and-slab formulation with the robust likelihood based on asymmetric Laplace distribution (ALD). The following merits have distinguished ssQLASSO from published studies. First, given a robust likelihood in the hierarchical formulation, the ssQLASSO is robust compared to ssLASSO, its major competitor, and other non-probust variable selection methods. Second, surprisingly, the choice of ALD as the likelihood has equipped the proposed ssQLASSO with an unexpected computational advantage to find the posterior mode, which is revealed by the soft-thresholding rule guided EM steps in the coordinate descent framework, just like the ssLASSO. This is a rare phenomenon since the soft-thresholding rule is generally not applicable in robust regularization with non-differentiable loss. Last but not least, the ssQLASSO has inherited the prominent properties of selective shrinkage and self-adaptivity to the sparsity pattern from ssLASSO, which has not been observed in any published robust variable selection methods. We have conducted extensive simulation studies under different model settings and a variety of heavy-tailed errors to demonstrate the utility of the proposed methods over alternatives. In the real data analysis, we have analyzed the TCGA LUAD and SKCM data under disease phenotypes with skewed distributions. The results show that the proposed method leads to biologically sensible findings with better predictive performance. The R package \emph{emBayes} is available at \url{https://cran.r-project.org/package=emBayes}.

\section{Statistical Methods}
\subsection{The Quantile Regression with Asymmetric Laplace Distribution}\label{sec:2.1}

To handle skewness in phenotypic traits and outliers, we considered using quantile regression below. Let $y_{i}$ be the response of the $i$-th subject $(1\leq i\leq n)$. We have $\boldsymbol z_{i}=(1,z_{i1},\dots,z_{iq})^{\top}$ being a $(q+1)$-dimensional vector of which the last $q$ components indicate clinical factors. In addition, $\boldsymbol x_{i}=(x_{i1},\dots,x_{ip})^{\top}$ represents a $p$-dimensional vector of genetic factors. The linear quantile regression model for the $\tau$-th $(0<\tau<1)$ quantile level is 
\begin{equation}\label{qr}
y_i=\boldsymbol z_i^\top\bm{\alpha_{\tau}}+\boldsymbol x_i^\top\bm{\beta_{\tau}}+\epsilon_{i,\tau},
\end{equation}
where $\bm{\alpha_{\tau}}=(\alpha_{0\tau},\cdots,\alpha_{q\tau})^\top$ consists of the regression coefficients corresponding to intercept and clinical covariates, and $\bm{\beta_{\tau}}=(\beta_{1\tau},\cdots,\beta_{p\tau})^\top$ are the regression coefficients for genetic factors or other types of omics features. The random error $\epsilon_{i\tau}$'s are independent with the $\tau$-th quantile being $0$. For simplicity of notation, the subscript ``$\tau$" has been omitted hereafter. The estimation of regression coefficients $\bm{\alpha}$ and $\bm{\beta}$ can be achieved by minimizing the following loss function
\begin{equation*}
L(\bm{\alpha},\bm{\beta})=\sum_{i=1}^n\rho_\tau(y_i-\boldsymbol z_i^\top\bm{\alpha}-\boldsymbol x_i ^\top\bm{\beta}),
\end{equation*}
in which the check loss function is defined as $\rho_{\tau}(\epsilon_{i})=\epsilon_{i}\{\tau-I(\epsilon_{i}<0)\}$. 

In literature, the asymmetric Laplace distribution (ALD) plays a pivotal role in probabilistic approaches to quantile regression, including the maximum likelihood estimation and Bayesian quantile regression (Yu and Moyeed, 2001)\cite{yu2001bayesian}. Specifically, with a fixed quantile level $\tau$ and scale parameter $\sigma(>0)$ that controls the skewness of the distribution , the probability density function of ALD can be expressed as 

\begin{equation*}
f(y_i|\mu_i,\sigma,\tau)=\frac{\tau(1-\tau)}{\sigma}\text{exp}\left( -\rho_{\tau}\left(\frac{y_i-\mu_i}{\sigma} \right)\right),
\end{equation*}
where $\mu_i=\boldsymbol z_i^\top\bm{\alpha}+\boldsymbol x_i^\top\bm{\beta}$. Then the joint distribution of $\boldsymbol Y=(y_1,...,y_n)^\top$ given $\boldsymbol x=(\boldsymbol x_1,...,\boldsymbol x_n)^\top$ and $\boldsymbol z=(\boldsymbol z_1,...,\boldsymbol z_n)^\top$ can be written as 

\begin{equation*}
f(\boldsymbol Y|\boldsymbol x,\boldsymbol z,\boldsymbol\beta,\boldsymbol\alpha,\sigma)=\tau^n(1-\tau)^n\sigma^{-n} \text{exp}\left( -\sum_{i=1}^{n}\rho_{\tau}\left(\frac{y_i-\mu_i}{\sigma} \right)\right).
\end{equation*}
Therefore, minimizing $L(\boldsymbol\alpha,\boldsymbol\beta)$ is equivalent to maximizing the (log-)likelihood function. The importance of asymmetric Laplace distribution to quantile regression has been further acknowledged since Kozumi and Kobayashi (2009) \cite{kozumi2011gibbs} have demonstrated that it is equivalent to the mixture of a scaled normal distribution and an exponential distribution, which significantly facilitates the formulation of the Bayesian hierarchical model. Define
\begin{equation*}
\zeta_1=\frac{1-2\tau}{\tau(1-\tau)}\ \text{and}\ \zeta_2=\sqrt{\frac{2}{\tau(1-\tau)}}.
\end{equation*}
Then the response variable has the form:
\begin{equation*}
y_i=\mu_i+\zeta_1 v_i+\zeta_2\sqrt{\sigma v_i}u_i,
\end{equation*}
where $v_i$ and $u_i$ follow exponential distribution with a scale parameter $\sigma(>0)$, $\text{Exp}(\sigma)$, and the standard normal distribution, $N(0,1)$, respectively. The hierarchical representation of $y_i$ following $ALD(\mu_i,\sigma, \tau)$ has the form as follows:	
\begin{equation*}
\begin{split}
y_i|v_i &\overset{ind}{\sim} \text{N}(\mu_i+\zeta_1v_i,\zeta_2^2\sigma v_i),\\
v_i &\overset{ind}{\sim} \text{Exp}(\sigma), \quad (i=1,...,n), \\
u_i &\overset{ind}{\sim} \text{N}(0,1), \quad (i=1,...,n).
\end{split}
\end{equation*}

\subsection{The Spike-and-Slab Quantile LASSO}

Identifying important genetic factors that are associated with the long-tailed disease traits demands robust variable selection \cite{wu2015selective,wu2019robust}. In the frequentist framework, adding the L1 norm of regression coefficients as a regularizer to the quantile regression model (\ref{qr}) leads to the following quantile-LASSO model:

\begin{equation}\label{equr:qlasso}
\underset{\boldsymbol\alpha,\boldsymbol\beta}{\text{min}}\sum_{i=1}^{n}\rho_{\tau}(y_i-\boldsymbol z_i^\top\bm{\alpha}-\boldsymbol x_i ^\top\bm{\beta})+\lambda \sum_{j=1}^{p}|\beta_j|,
\end{equation}
where $\lambda>0$ is the tuning parameter determining the model sparsity. At the 50$\%$ quantile level, model (\ref{equr:qlasso}) reduces to the least absolute deviation (LAD) LASSO \cite{wang2007robust}. On the other hand, with the aforementioned formulation of quantile regression using asymmetric Laplace distribution, Li et al. \cite{li2010bayesian} have developed the Bayesian quantile LASSO by imposing the i.i.d conditional Laplace prior on regression coefficients corresponding to high-dimensional predictors \cite{park2008bayesian,casella2010penalized}:
\begin{equation}\label{equr:laplace}
\beta_j|s \overset{ind}{\sim} \frac{1}{2s}\text{exp}\left(-\frac{|\beta_j|}{s}\right),
\end{equation}
where $s=\frac{\sigma}{\lambda}$. Li et al \cite{li2010bayesian} have further demonstrated that the quantile LASSO estimator can be viewed as a posterior mode of the regression coefficients under such a prior specification.


We consider the following spike-and-slab prior \cite{rovckova2018spike,tang2017spike,tang2017spikeGLM}:
\begin{equation}\label{equr:ssprior}
\beta_j|\gamma_j,s_0,s_1 \overset{ind}{\sim} (1-\gamma_j)\pi(\beta_j|s_0)+\gamma_j\pi(\beta_j|s_1),
\end{equation}
where the spike component $\pi(\beta_j|s_0)$ and slab component $\pi(\beta_j|s_1)$ are defined using the conditional Laplace prior in (\ref{equr:laplace}) with scale parameters $s_1>s_0>0$. When the indicator $\gamma_j=1$, the regression coefficient $\beta_j$ is assumed to have coefficients with larger magnitude and be modeled by the slab part. Then the corresponding genetic factor is associated with the disease trait. Otherwise, $\gamma_j$ being 0 indicates that $\beta_j$ is small and close-to-0, following the spike distribution. Therefore, the corresponding genetic factor does not contribute to the variation of the phenotype. Alternatively, the spike-and-slab formulation can be written as   
$\beta_j|\gamma_j,S_j \overset{ind}{\sim} \pi(\beta_j|S_j)=\frac{1}{2S_j}\text{exp}\left( -\frac{|\beta_j|}{S_j}\right)$, where $S_j=(1-\gamma_j)s_0+\gamma_js_1$.

In order to integrate the frequentist Bayesian quantile LASSO with the Bayesian quantile spike-and-slab regression for the proposed ssQLASSO, we first specify the following hierarchical representation. We assign the independent Bernoulli prior on $\gamma_j$, ($j=1,...,p$):
\begin{equation*}
\gamma_j|\theta \overset{ind}{\sim} \text{Ber}(\theta)= \theta^{\gamma_j}(1-\theta)^{1-\gamma_j}, 
\end{equation*}
with the parameter $\theta$ having a uniform prior $\theta \sim U(0,1)$. After $\theta$ being treated as unknown and random, it can be viewed as a global shrinkage parameter that equals $p(\gamma_j=1|\theta)$, the prior probability of nonzero coefficients in $\bm{\beta}$. The prior expectation of $S_j$, $E(S_j)=(1-\theta)s_0+\theta s_1$, suggests that the shrinkage is imposed adaptively on each coordinate of $\bm{\beta}$.

Besides, the regression coefficients for intercept and clinical factors, $\alpha_k$, are assumed to have independent normal priors with mean 0 and variance $V_k$:
\begin{equation*}
\alpha_k\overset{ind}{\sim}\text{N}(0,V_k).
\end{equation*}
And the scale parameter $\sigma>0$ has inverse-gamma prior distribution with $a>0$ and $b>0$ being constants:
\begin{equation*}
\sigma\sim \text{IG}(a,b)=\sigma^{-a-1}\text{exp}\left( -\frac{b}{\sigma}\right).
\end{equation*}
For computational convenience, we choose $V_k=10^3$ and $a=b=1$. More discussions on the sensitivity analysis will be provided in the Section of Simulation.

\textbf{Remark}: 
Together with the ALD likelihood, the hierarchical model with the prior in (\ref{equr:ssprior}) leads to the Bayesian quantile LASSO proposed in Li et al. \cite{li2010bayesian} when the two scale parameters are equal, i.e $s_0=s_1$. A well recognized limitation of Bayesian quantile LASSO is that the posterior estimates of $\bm{\beta}$ lack exact sparsity, and its frequentist counterpart, the quantile LASSO, incurs estimation bias by over-shrinking coefficients with large magnitude. 

The ssQLASSO enjoys the selective shrinkage and self-adaptive properties that are inherited from the ssLASSO \cite{rovckova2018spike} to facilitate more accurate and robust identification of important cancer genomcics features. The selective shrinkage can be achieved through a small amount of penalty on large $|\beta_j|$, and a large amount of shrinkage on small $|\beta_j|$. Meanwhile, the self-adaptivity ensures stabilizing coefficients with large magnitude even when the tuning is at its extremes. Therefore regularization of ssQLASSO is essentially different from that of the adaptive (quantile) LASSO \cite{zou2006adaptive}. In addition, compared to Bayesian quantile LASSO \cite{li2010bayesian}, its prior structure of adaptively mixing the two Laplace distributions leads to the modal estimate of $\bm{\beta}$ with exact sparsity. When $s_0 \rightarrow 0$, the prior specified by (\ref{equr:ssprior}) reduces to the point mass spike-and-slab priors.

\subsection{Implementation via the EM Algorithm}

The elicitation of the continuous spike-and-slab mixture double exponential prior in ssQLASSO is distinct from the point mass spike-and-slab priors in Bayesian regularized quantile regression mainly in that the scale parameters $s_0$ and $s_1$ in (\ref{equr:ssprior}) are treated as deterministic, instead of being random and assigned with conjugate priors  \cite{ren2023robust,zhou2023bayesian}. Such a difference has paved the way for a extremely fast EM algorithm based on the coordinate descent idea to find the modal estimates of $\bm \beta$ with exact sparsity, and freed us from using the computationally intensive MCMC algorithm in the fully robust Bayesian analysis. 

The $\boldsymbol v=(v_1,\dots,v_n)^\top$ from the probabilistic formulation of quantile regression and indicator variables $\boldsymbol \gamma=(\gamma_1,\dots,\gamma_p)^\top$ are treated as missing values in the proposed EM algorithm. Recall that  $S_j=(1-\gamma_j)s_0+\gamma_j s_1$, which also has missingness due to $\gamma_j$. Subsequently, at fixed quantile level $\tau$, let $\bm{\phi}=(\bm{\alpha},\bm{\beta},\sigma,\boldsymbol \gamma,\boldsymbol v,\theta)$ the log joint posterior density is given as:

\begin{equation}\label{equr:lBQLSS}
\begin{split}
Q&=\text{log}p(\bm{\alpha},\bm{\beta},\sigma,\boldsymbol \gamma,\boldsymbol v,\theta|\boldsymbol Y)\\
&=\text{log}p(\boldsymbol Y|\bm{\alpha},\bm{\beta},\sigma,\boldsymbol v)+\sum_{k=0}^q\text{log}p(\alpha_k)+\sum_{j=1}^{p}\text{log} p(\beta_j|S_j)+\text{log}p(\sigma)+\sum_{i=1}^{n}p(v_i)\\
&+\sum_{j=1}^{p}\text{log} p(\gamma_j|\theta)+\text{log} p(\theta)\\
&\propto -\frac{3n}{2}\text{log}(\sigma)-\frac{1}{2}\sum_{i=1}^{n}\text{log}(v_i)-\sum_{i=1}^{n}\left(\frac{(y_i-\boldsymbol z_i^\top\bm{\alpha}-\boldsymbol x_i^\top\bm{\beta})^2}{2\zeta_2^2\sigma}v_i^{-1}-\frac{\zeta_1(y_i-\boldsymbol z_i^\top\bm{\alpha}-\boldsymbol x_i^\top\bm{\beta})}{\zeta_2^2\sigma}+\frac{\zeta_1^2}{2\zeta_2^2\sigma}v_i\right)\\
&\ \ \ -\sum_{k=0}^q\frac{\alpha_k^2}{2V_k}-\sum_{j=1}^{p}\frac{1}{S_j}|\beta_j |-\frac{b}{\sigma}+(a+1)\text{log}(\sigma)-\frac{1}{\sigma}\sum_{i=1}^{n}v_i+\sum_{j=1}^{p}\left[\gamma_j \text{log}\theta+(1-\gamma_j)\text{log}(1-\theta)\right].
\end{split}
\end{equation}
As terms involving $v_i$ ($i=1,...,n$) and $\gamma_j$ ($j=1,...,p$) are treated as missing, the EM algorithm estimates $(\bm{\alpha},\bm{\beta},\sigma,\theta)$ through averaging the missing values across the parameters' respective posterior distributions. At E-step, we calculate the conditional posterior expectations of $v_i^{-1}$, $v_i$, $\gamma_j$ and $S_j^{-1}$. In (\ref{equr:lBQLSS}), all the terms that involve $v_i$ can be collectively written as:
\begin{equation}\label{GIGv}
v_i^{-1/2}\text{exp}\left(\frac{1}{2}(w_{1i}^2v_i^{-1}+w_2^2v_i) \right),
\end{equation}
where $w_{1i}^2=\frac{(y_i-\boldsymbol z_i^\top\bm{\alpha}-\boldsymbol x_i^\top\bm{\beta})^2}{\zeta_2^2\sigma}$ and $w_{2}^2=\frac{2}{\sigma}+\frac{\zeta_1^2}{\zeta_2^2\sigma}$, which is the kernel of a generalized inverse Gaussian (GIG) distribution denoted as $v_i|y_i,\bm{\alpha},\bm{\beta},\sigma \sim GIG\left( \frac{1}{2},w_{1i},w_{2} \right).$  
Here, the probability density function of a random variable $\Omega$ following the generalized inverse Gaussian (GIG) distribution, $GIG(t,c_1,c_2)$, is given as
\begin{equation*}
f(\Omega|t,c_1,c_2)=\frac{(c_2/c_1)^{t}}{2K_{t}(c_1c_2)}\Omega^{t-1}\text{exp}\left(-\frac{1}{2}(c_1^2\Omega^{-1}+c_2^2\Omega) \right),
\end{equation*}
where $\Omega>0$, $c_1>0$, $c_2>0$ and the real number $t$ is a parameter used in function $K(\cdot)$, the modified Bessel function of the third kind (Karlis, 2002)\cite{karlis2002type}. The moments of $\Omega$ are given by
\begin{equation*}
E(\Omega^{\kappa})=\left( \frac{c_{1}}{c_2} \right)^{\kappa}\frac{K_{t+\kappa}(c_{1} c_2)}{K_{t}(c_{1} c_2)},\kappa\in \text{R}.
\end{equation*}
When choosing $t=\frac{1}{2}$ in the above pdf under (\ref{GIGv}), we have
\begin{equation}\label{vi}
\begin{split}
&v_{ni}^{(d)}=E\left((v_i^{-1})^{(d)}\right)=\left( \frac{w_{1i}^{(d)}}{w_2^{(d)}} \right)^{-1}\frac{K_{-1/2}(w_{1i}^{(d)} w_2^{(d)})}{K_{1/2}(w_{1i}^{(d)} w_2^{(d)})},\\
&v_{pi}^{(d)}=E\left(v_i^{(d)}\right)=\left( \frac{w_{1i}^{(d)}}{w_2^{(d)}} \right)\frac{K_{3/2}(w_{1i}^{(d)} w_2^{(d)})}{K_{1/2}(w_{1i}^{(d)} w_2^{(d)})}.
\end{split}
\end{equation}
As for $\gamma_j$, its conditional posterior expectation at the $d$-th iteration can be derived as:
\begin{equation}\label{gammaj}
\begin{split}
\eta_j^{(d)}&=p(\gamma_j=1|\beta_j^{(d)},\theta^{(d)},\boldsymbol Y)\\
&=\frac{p(\beta_j^{(d)}|\gamma_j=1,s_1)p(\gamma_j=1|\theta^{(d)})}{p(\beta_j^{(d)}|\gamma_j=0,s_0)p(\gamma_j=0|\theta^{(d)})+p(\beta_j^{(d)}|\gamma_j=1,s_1)p(\gamma_j=1|\theta^{(d)})}\\
&=\frac{\pi(\beta_j^{(d)}|s_1)\theta^{(d)}}{\pi(\beta_j^{(d)}|s_0)(1-\theta^{(d)})+\pi(\beta_j^{(d)}|s_1)\theta^{(d)}},
\end{split}
\end{equation}
where $\pi(\beta_j^{(d)}|s_0)=\frac{1}{2s_1}\text{exp}\left(-\frac{|\beta_j^{(d)}|}{s_0}\right)$ and $\pi(\beta_j^{(d)}|s_1)=\frac{1}{2s_1}\text{exp}\left(-\frac{|\beta_j^{(d)}|}{s_1}\right)$. Consequently, the conditional expectation of $(S_j^{-1})^{(d)}$ can be computed as:
\begin{equation}\label{Sj}
(\tilde{S}_j^{-1})^{(d)}=E((S_j^{-1})^{(d)}|\beta_j^{(d)})=E\left(\frac{1}{(1-\gamma_j)s_0+\gamma_j s_1}\middle|\beta_j^{(d)}\right)=\frac{1-\eta_j^{(d)}}{s_0}+\frac{\eta_j^{(d)}}{s_1}.
\end{equation}
At the M-step, we update $\bm{\alpha}^{(d+1)}$, $\bm{\beta}^{(d+1)}$, $\sigma^{(d+1)}$ and $\theta^{(d+1)}$ by maximizing the log joint posterior density at $d$-th iteration, or $Q(\bm{\phi}|\bm{\phi}^{(d)})$. The coefficient vector $\bm{\alpha}$ and $\bm{\beta}$ are updated in a component-wise manner using the cyclic coordinate decent algorithm \cite{friedman2010regularization}. By solving $\frac{\partial Q(\bm{\phi}|\bm{\phi}^{(d)})}{\partial \alpha_l}=0$, $\frac{\partial Q(\bm{\phi}|\bm{\phi}^{(d)})}{\partial \beta_m}=0$, $\frac{\partial Q(\bm{\phi}|\bm{\phi}^{(d)})}{\partial \sigma}=0$ and $\frac{\partial Q(\bm{\phi}|\bm{\phi}^{(d)})}{\partial \theta}=0$, where $l=1,...,q$ and $m=1,...,p$, we obtain the following expressions
\begin{equation}\label{cd}
\begin{split}
\theta^{(d+1)}&=\frac{1}{p}\sum_{j=1}^{p}\eta_j^{(d)},\\
\sigma^{(d+1)}&=\frac{\sum_{i=1}^{n}\left(v_{ni}^{(d)}(y_i-\boldsymbol x_i^\top\bm{\beta}^{(d)}-\boldsymbol z_i^\top\bm{\alpha}^{(d)})^2-2(y_i-\boldsymbol x_i^\top\bm{\beta}^{(d)}-\boldsymbol z_i^\top\bm{\alpha}^{(d)})\zeta_1+v_{pi}^{(d)}(\zeta_1^2+2\zeta_2^2) \right)+b}{(3n+2a+2)\zeta_2^2},\\
\alpha_l^{(d+1)}&=\frac{\sum_{i=1}^{n}v_{ni}^{(d)}\left( y_i-\boldsymbol x_i^\top\bm{\beta}^{(d)}-\sum_{k=1}^{l-1}z_{ik}\alpha_k^{(d+1)}-\sum_{k=l+1}^{q}z_{ik}\alpha_k^{(d)}\right)z_{il}-\zeta_1\sum_{i=1}^nz_{il}}{\frac{\zeta_2^2\sigma^{(d+1)}}{V_l}+\sum_{i=1}^nv_{ni}^{(d)}z_{il}^2},\\
\beta_m^{(d+1)}&=\frac{\text{sgn}(T_m^{(d+1)})\left(\left| T_m^{(d+1)}\right|-(\tilde{S}_m^{-1})^{(d)}\right)_+}{\sum_{i=1}^{n}\frac{v_{ni}^{(d)}x_{im}^2}{\zeta_2^2\sigma^{(d+1)}}},
\end{split}
\end{equation}
where
\begin{equation*}
T_m^{(d+1)}=\sum_{i=1}^{n}\frac{v_{ni}^{(d)}\left(y_i-\sum_{j=1}^{m-1}x_{ij}\beta_j^{(d+1)}-\sum_{j=m+1}^{p}x_{ij}\beta_j^{(d)}-\boldsymbol z_i^\top\bm{\alpha}^{(d+1)} \right)x_{im}}{\zeta_2^2\sigma^{(d+1)}}-\sum_{i=1}^{n}\frac{\zeta_1}{\zeta_2^2\sigma^{(d+1)}}x_{im}.
\end{equation*}

The ssQLASSO modal estimator on $\beta_m$, the regression coefficient corresponding to the $m$th genetic factor, shares a remarkable resemblance to the LASSO soft thresholding rule  \cite{tibshirani1996regression,friedman2010regularization}. This phenomena is rarely observed in robust penalization with non-differentiable loss functions because thresholding rule exists under smooth losses. In high-dimensional quantile regression, more computationally challenging algorithms, including linear programming and weighted median regression among others, are required to accommodate the non-differentiability in optimization \cite{wu2009variable,peng2015iterative,li20081,ren2019robust}. Therefore, ssQLASSO is computationally more appealing over the freuquentist regularized quantile regression because it can naturally fit into the extremely fast coordinate descent algorithm that merely has a complexity of $O(np)$ when updating $\boldsymbol{\beta}$.

The update on $\beta_m$ from (\ref{cd}) further justifies the selective shrinkage property discussed in Section 2.2. Rather than a fixed tuning parameter for all coordinates of $\bm{\beta}$, $\tilde{S}_m^{-1}$ is coordinate-specific in that the threshold is smaller for larger $|\beta_m|$, leading to different amount of shrinkage for $|\beta_m|$, ($m=1,...,p$), to avoid the estimation bias in quantile LASSO. In addition, the threshold $\tilde{S}_m^{-1}$ in ssQLASSO can eventually keep large $|\beta_m|$ stable via slight shrinkage even when the tuning parameter $s_0$ is at its extremes, i.e. close to 0. Therefore the ssQLASSO is self-adaptive to the underlying sparsity pattern, and is essentially different from adaptive (quantile) LASSO which allocate fixed coordinate-specific weight in the penalty for $\beta_m$ so all coefficients vanish when the tuning parameter, proportional to $1/s_0$ as shown in (\ref{equr:laplace}), is at infinity. Note that the intercept and clinical covariates are not subject to selection, so the update of regression coefficients $\bm \alpha$ in (\ref{cd}) does not follow the soft-thresholding rule.   

In published studies, the Schwarz Information Criterion (SIC) has been widely adopted for tuning selection in high-dimensional quantile regression \cite{wang2009quantile,tang2013variable}, which is defined as
\begin{equation*}
SIC=\text{log}\sum_{i=1}^nL(y_i-\hat{y_i})+\frac{\text{log}n}{2n}edf,
\end{equation*}
where $L(\cdot)$ is the quantile check loss function, and $edf$ is the number of zero residuals. We adopt SIC  to choose the optimal pair of tuning parameters ($s_0,s_1$). More detailed analyses of tuning selection are provided in the Section of Simulation. Under the fixed tuning parameters, the algorithm proceeds following steps outlined in Table \ref{alg2}:





\begin{table} [hbt!]
	\def\arraystretch{1.3}
	\begin{center}
		\caption{EM algorithm for the spike-and-slab quantile LASSO.}\label{alg2}
		\centering
		\fontsize{10}{10}\selectfont{
			\begin{tabularx}{\textwidth}{ m{0.9\textwidth} }
				\hline
				\vspace{0.05cm}
				\textbf{Algorithm} \\[0.1ex]
				\hline
				\vspace{0.1cm}
				\hspace{1em} Initialize iteration $d = 0$, quantile $\tau$, coefficients $\bm{\alpha}^{(0)}$, $\bm{\beta}^{(0)}$, $\theta^{(0)}$ and $\sigma^{(0)}$;\\
				\hspace{1em} \textbf{Repeat} \\
				\hspace{3em} \textbf{E-step:}\\
				\hspace{3em} Compute the conditional expectation of $(v_i^{-1})^{(d)}$, $v_i^{(d)}$, $\gamma_j^{(d)}$, $(\tilde{S}_j^{-1})^{(d)}$ via (\ref{vi}), (\ref{gammaj}), (\ref{Sj});\\
				\hspace{3em} \textbf{M-step:}\\
				\hspace{3em} Update $\theta^{(d+1)}$, $\sigma^{(d+1)}$, $\alpha_l^{(d+1)}$ and $\beta_m^{(d+1)}$ through (\ref{cd}); \\
				\hspace{3em} Calculate the updated log joint posterior density and check if $|Q^{(d+1)}-Q^{(d)}|<\delta$;\\
				\hspace{3em} $d\leftarrow d +1$; \\
				\hspace{1em} \textbf{until} Convergence ($|Q^{(d+1)}-Q^{(d)}|<\delta$ for a small value $\delta>0$).\\
				\hline
		\end{tabularx} }
	\end{center}
	\centering
\end{table}
\FloatBarrier

\section{Simulation Study}
We have conducted simulation studies to demonstrate the effectiveness of the proposed spike-and-slab quantile LASSO (ssQLASSO). Three alternatives, quantile LASSO (QLASSO), spike-and-slab LASSO (ssLASSO) \cite{rovckova2018spike,tang2017spike}, and LASSO have been included for comparison. Details of the EM algorithm to fit ssLASSO are provided in the Appendix. R packages  \emph{glmnet} \cite{friedman2010regularization} and \emph{rqPen} \cite{sherwood2023package} have been adopted to fit the LASSO and QLASSO, respectively. For all the alternative methods, the tuning parameters have been chosen via cross-validation under the squared error for ssLASSO and LASSO, and the check loss for QLASSO. Specifically, we have adopted the function \emph{rq.pen.cv} from package \emph{rqPen} to choose tuning parameter in QLASSO.

The data are generated following model (\ref{qr}) with sample size $n=400$. The dimensions are $p=800$ and $p=1600$ for genetic factors $\boldsymbol x$ which are simulated from multivariate normal distributions with (1) the auto-regression (AR) structure where the $i$th and $j$th genes have correlation $\rho^{|i-j|}$, and (2) banded structure where  the $i$th and $j$th genes have correlation $\rho$ when $|i-j|=1$ and $0$ otherwise. So the diagonal entries in both structures are 1's. The parameter $\rho$ in both cases is set to 0.5. In model (\ref{qr}), the clinical covariates $\boldsymbol z$ are not subject to selection, so without loss of generality, we set corresponding coefficients to 0, and the intercept $\alpha_0$ to 2. 15 nonzero components from the coefficient vector $\boldsymbol \beta$ are generated from uniform distribution $U[0.6,0.8]$, with the rest being 0s. Five probability distributions are considered for random error $\epsilon_i$ in model (\ref{qr}):  $N$($\mu$, 1)(Error 1),  $t$(2) with mean $\mu$(Error 2), LogNormal($\mu$,1) (Error 3), 80\%$N$($\mu$,1) + 20\%$N$($\mu$, 3) (Error 4) and Laplace($\mu$,1) (Error 5). All the errors have long-tailed distributions except Error 1. For each error distribution, the location parameter $\mu$ has been adjusted so the $\tau$th quantile of the distribution is 0. In addition to the homogeneous random error in model (\ref{qr}), we have also considered the heterogeneous model:    
\begin{equation}\label{heter}
y_i=2+\boldsymbol x_i^\top\bm{\beta}+(1+x_{i2})\epsilon_i,
\end{equation}
where $\boldsymbol \beta$ is generated the same as the coefficient vector under the homogeneous model except that $\beta_2$, the coefficient corresponding to the 2nd genetic factor, is always nonzero. The error term becomes non-$i.i.d.$ in this case.

Therefore there are 8 case scenarios given different dimensionality ($p$=800 and 1600), correlation structure (AR1 and banded) and model types (heterogeneous and homogeneous). For each setting, the comparison has been conducted under 3 quantile levels (30$\%$,50$\%$ and 70$\%$) and 5 error distributions. The number of true positives (TP), false positives (FP), $\text{F}_1$ score (F1) and Matthews correlation coefficient (MCC) have been adopted to evaluate identification performance. In terms of estimation accuracy, we compute $\sum|\hat{\beta_j}-\beta_j|$, the L1 difference between the estimated and true regression coefficients. Additional simulations have been provided at the end of this section.

Table \ref{sim2} shows the identification and estimation results under the homogeneous model with AR-1 correlation and $(n,p)$= (400,1600) across quantile levels $\tau=30\%$, $50\%$ and $70\%$, and 5 error distributions. The advantage of ssQLASSO over alternatives can be convincingly shown when the model errors are heavy-tailed. For example, under $t(2)$ error and quantile level $\tau=30\%$ , it can be observed that ssQLASSO almost identifies all the important signals with a TP of 14.63 (sd0.81) and nearly negligible number of FPs, 0.33 (sd0.84). Although the number of TPs detected by QLASSO is comparable, the FPs, 34.67 (sd5.57), are much higher. The two non-robust approaches are completely dominated by ssQLASSO. In particular, ssLASSO can only discover roughly $60\%$ of true effects with the TP of 9.33 (sd2.75), and LASSO leads to the largest FPs of 70.47 (sd29.01). In terms of both metrics of F1 score and MCC that more comprehensively reflect the identification performance, the proposed ssQLASSO has also demonstrated the best performance. In particular, the F1 score of 0.98 (sd0.04) and MCC of 0.98 (sd0.04) are the largest among the measures of all methods under comparison. The superior performance of ssQLASSO in identification also results in the smallest estimation error of 2.81 (sd1.12) in contrast to the alternatives. The merit of ssQLASSO can be further established under the rest of skewed model errors (Error3--5) at all quantile levels in Table \ref{sim2}. On the other hand, with the light-tailed $N$(0,1) error, ssQLASSO is the 2nd best method which is marginally outperformed by its non-robust counterpart, ssLASSO. For instance, at $\tau=30\%$, the F1 score and MCC of ssQLASSO are both 0.98 (sd0.03), with an estimation error of 0.69 (sd0.20), while those metrics for ssQLASSO are 0.97 (sd0.04), 0.97 (sd0.04) and 0.87 (sd0.17), respectively. This phenomenon indicates that although the ssQLASSO has been proposed as a robust method to accommodate long-tailed errors and outlying observations , it is still very competitive under the $N(0,1)$ error.     

\begin{figure}[H]
\begin{center}
		\includegraphics[height=0.35\textheight]{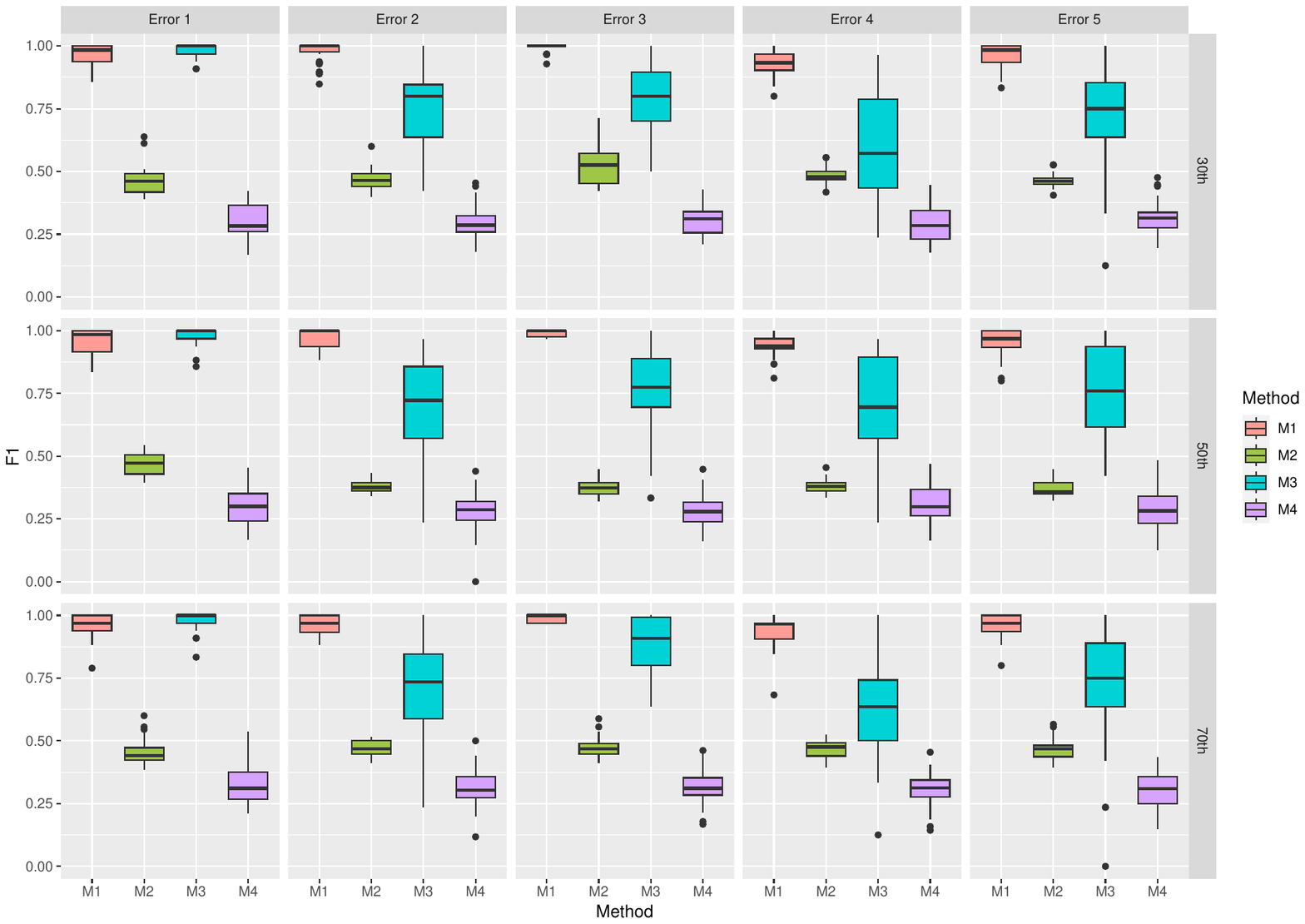}
		\caption{\footnotesize  F1 score under AR-1 correlation, homogeneous error and $(n,p)=(400,1600)$. Methods M1-M4 correspond to ssQLASSO, QLASSO, ssLASSO and LASSO, respectively. }\label{r1}
		\end{center}
	
	\begin{center}
		\includegraphics[height=0.35\textheight]{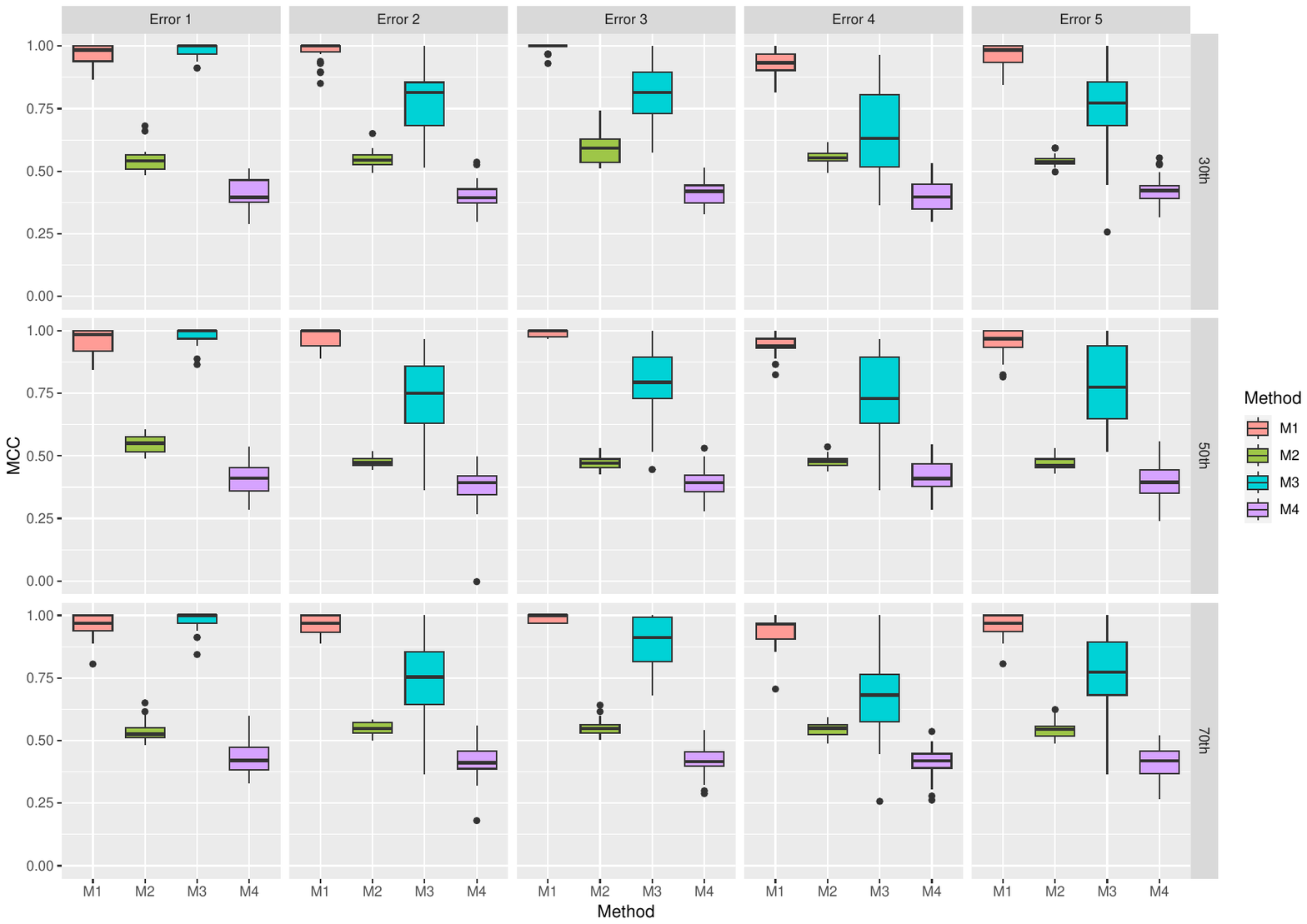}
		\caption{\footnotesize MCC score under AR-1 correlation, homogeneous error and $(n,p)=(400,1600)$. Methods M1-M4 correspond to ssQLASSO, QLASSO, ssLASSO and LASSO, respectively.}\label{r2}
		\end{center}
	\end{figure}

\begin{figure}[H]
	\begin{center}
		\includegraphics[height=0.35\textheight]{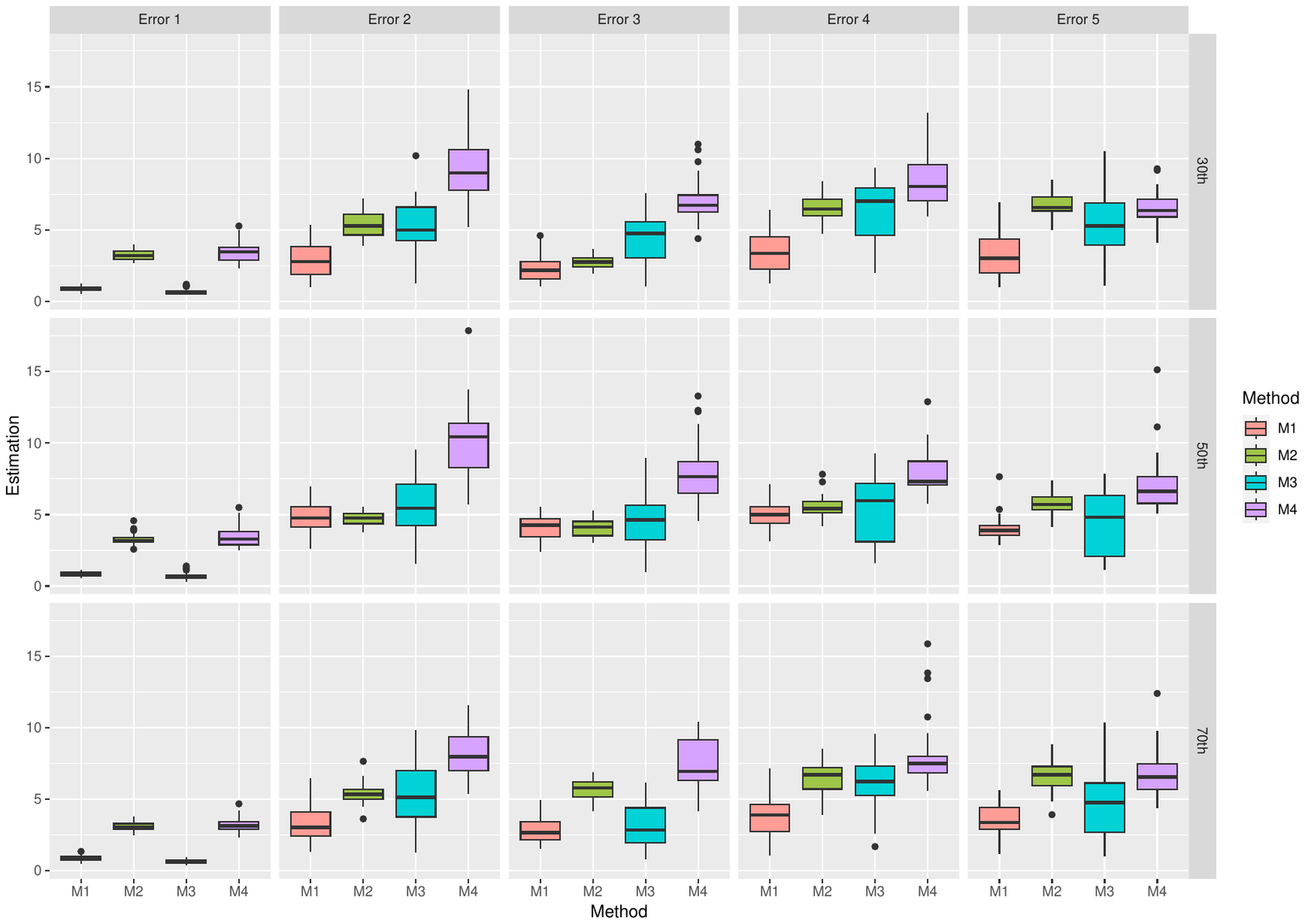}
		\caption{\footnotesize Estimation error under AR-1 correlation, homogeneous error and $(n,p)=(400,1600)$. Methods M1-M4 correspond to ssQLASSO, QLASSO, ssLASSO and LASSO, respectively.}\label{r3}
		\end{center}
	\end{figure}

In addition to tabulating the mean and standard deviation of the measures assessing identification and estimation performance in Table \ref{sim2}, we have graphically displayed the box-plots of $\text{F}_1$ score, MCC and estimation error over 100 replications in Figure \ref{r1}, \ref{r2} and \ref{r3}, respectively, which visualize Table \ref{sim2} from different perspectives.     


Figure \ref{r1} shows the box-plots of F1 scores over 100 replicates across 5 model errors and 3 quantile levels. Under the $N$(0,1) error, the box-plots of ssQLASSO and ssLASSO are close to the top of all the sub-panels, indicating majority of the F1 scores computed based on the signals identified by the two methods are close to 1. Their performance are comparable although ssLASSO is slightly better, as shown by a smaller interquartile range (IQR) of F1 scores. Besides, the box-plots of QLASSO and LASSO are corresponding to much lower F1 scores, which further demonstrates the advantage of the spike-and-slab (quantile) LASSO prior in identification accuracy. As the model errors become heavy-tailed in the next 4 columns, ssQLASSO has consistently outperformed the non-robust ssLASSO with most of its F1 scores centering around 1, except a few outliers. Interestingly, for all methods under comparison, ssLASSO results in box-plots with the largest IQR of F1 scores in the presence of data heterogeneity and outliers, revealing the instability of ssLASSO and further justifying the pressing need to develop the robust ssQLASSO. In addition to F1 score, MCC has also been frequently used as a criterion to assess identification usually providing more accurate results in cases of imbalanced data, which is applicable in the simulation with sparsity assumption, e.g., only 15 important features out of a total of 1600 gene expressions. By comparing Figure \ref{r2} to Figure \ref{r1}, we can draw similar conclusions on the superiority of the proposed ssQLASSO in terms of identification. 

Figure \ref{r3} shares a similar layout of Figure \ref{r1} and \ref{r2}. The only difference is that the box-plots in Figure \ref{r3} represent estimation error, $\sum|\hat{\beta_j}-\beta_j|$, across 100 replications. We can observe that ssQLASSO has superior prediction performance especially under heavy-tailed errors, as indicated by the lower box-plots with small IQRs, in spite of a few outlying estimation errors. Meanwhile, ssLASSO appears relatively unstable with much larger IQRs of the box-plots under all the error distributions except Error 1, $N$(0,1). The strong contrast between the two additionally justifies the merit of ssQLASSO over its non-robust counterpart to handle data contamination.    

In the appendix, we have presented results from simulations under the rest 7 out of the 8 case scenarios based on models with different error ( $i.i.d$ and non-$i.i.d$), dimensionality ($p$=800 and 1600), and correlations (AR1 and banded). Under the homogeneous model with the $i.i.d$ error,  Table \ref{sim1} shows the performance of all 4 methods with AR-1 correlation and $p$=800, while Table \ref{sim3} and \ref{sim4} provide the numeric summaries of identification and estimation performance of banded correlation with dimension 800 and 1600, respectively. We can observe patterns similar to those in Table \ref{sim2}, and conclude the advantage of the proposed ssQLASSO. All the methods have also been examined when the data are simulated using the heterogeneous model (\ref{heter}). With the non-$i.i.d$ model errors, Table \ref{sim5}--\ref{sim8} in the appendix list numeric measures in identification and estimation from settings according to different combination of correlation (AR1 and banded) and number of predictors ($p$=800 and 1600). The superiority of ssQLASSO over the alternatives can be drawn not merely under Error2--5. Note that even using $N$(0,1), the random errors are no longer independent in model (\ref{heter}). So it is not surprising to observe the better performance of ssQLASSO under Error 1 over non-robust ones. For instance, Table \ref{sim2} and \ref{sim6} are corresponding to the settings that differ only in the type of error under the same correlation (AR-1) and dimension ($p$=1600). With quantile level $\tau$=0.3 and Error1, the estimation errors of ssQLASSO and ssLASSO in Table \ref{sim2} are 0.87 (sd0.17) and 0.69 (sd0.20), respectively. As the errors become non-$i.i.d$ in Table \ref{sim6}, ssQLASSO has a smaller L1 error of 1.07 (sd0.18), compared to 1.26 (sd0.20) obtained from ssLASSO.

\begin{table} [H]
	\def\arraystretch{1.5}
	\captionsetup{font=scriptsize}
	\begin{center}
		\caption{Evaluation of homogeneous model under AR-1 correlation, ($n,p$)=(400,1600) with 100 replicates.}\label{sim2}
		\centering
		\fontsize{7}{7}\selectfont{
			\begin{tabular}{ l l l l l l l l }
				\hline
				$n=400$&$p=1600$&& \multicolumn{1}{c}{TP}& \multicolumn{1}{c}{FP} & \multicolumn{1}{c}{F1} & \multicolumn{1}{c}{MCC} & \multicolumn{1}{c}{Estimation}\\
				\hline
				
				$\tau=0.3$ & Error1 & ssQLASSO & 15.00(0.00) & 1.10(1.40) & 0.97(0.04) & 0.97(0.04) & 0.87(0.17) \\ 
   &  & QLASSO & 15.00(0.00) & 35.73(7.54) & 0.46(0.06) & 0.54(0.05) & 3.24(0.35) \\ 
   &  & ssLASSO & 15.00(0.00) & 0.70(0.92) & 0.98(0.03) & 0.98(0.03) & 0.69(0.20) \\ 
   &  & LASSO & 15.00(0.00) & 75.87(27.29) & 0.30(0.07) & 0.41(0.06) & 3.47(0.70) \\
				\cline{2-8}
				& Error2 & ssQLASSO & 14.63(0.81) & 0.33(0.84) & 0.98(0.04) & 0.98(0.04) & 2.81(1.12) \\ 
   &  & QLASSO & 15.00(0.00) & 34.67(5.57) & 0.47(0.04) & 0.55(0.03) & 5.41(1.00) \\ 
   &  & ssLASSO & 9.33(2.75) & 0.07(0.25) & 0.75(0.14) & 0.78(0.12) & 5.28(1.87) \\ 
   &  & LASSO & 13.93(3.25) & 70.47(29.01) & 0.29(0.09) & 0.40(0.06) & 9.40(2.63) \\
				\cline{2-8}
				& Error3 & ssQLASSO & 14.90(0.40) & 0.10(0.31) & 0.99(0.02) & 0.99(0.02) & 2.31(0.92) \\ 
   &  & QLASSO & 15.00(0.00) & 27.80(8.29) & 0.53(0.08) & 0.60(0.06) & 2.76(0.46) \\ 
   &  & ssLASSO & 10.20(3.14) & 0.07(0.25) & 0.79(0.15) & 0.81(0.13) & 4.47(1.88) \\ 
   &  & LASSO & 14.93(0.37) & 71.73(19.48) & 0.30(0.05) & 0.41(0.05) & 7.05(1.57) \\ 
				\cline{2-8}
				& Error4 & ssQLASSO & 14.10(1.21) & 1.13(1.31) & 0.93(0.05) & 0.93(0.05) & 3.48(1.40) \\ 
   &  & QLASSO & 14.83(0.38) & 31.57(4.41) & 0.49(0.04) & 0.56(0.03) & 6.55(0.92) \\ 
   &  & ssLASSO & 7.03(3.68) & 0.17(0.38) & 0.60(0.22) & 0.65(0.18) & 6.28(2.19) \\ 
   &  & LASSO & 15.00(0.00) & 79.73(29.60) & 0.29(0.08) & 0.40(0.07) & 8.41(1.95) \\ 
				\cline{2-8}
				& Error5 & ssQLASSO & 14.83(0.46) & 1.03(1.63) & 0.96(0.05) & 0.96(0.05) & 3.22(1.47) \\ 
   &  & QLASSO & 14.90(0.40) & 34.40(4.21) & 0.47(0.03) & 0.54(0.02) & 6.69(0.90) \\ 
   &  & ssLASSO & 8.90(3.92) & 0.27(0.58) & 0.70(0.22) & 0.74(0.19) & 5.47(2.54) \\ 
   &  & LASSO & 15.00(0.00) & 68.17(19.49) & 0.32(0.06) & 0.42(0.05) & 6.57(1.13) \\ 
				\hline
				
				$\tau=0.5$& Error1 & ssQLASSO & 15.00(0.00) & 1.33(1.75) & 0.96(0.05) & 0.96(0.05) & 0.85(0.17) \\ 
   &  & QLASSO & 15.00(0.00) & 34.83(6.44) & 0.47(0.05) & 0.55(0.04) & 3.28(0.40) \\ 
   &  & ssLASSO & 15.00(0.00) & 0.87(1.25) & 0.97(0.04) & 0.97(0.04) & 0.70(0.24) \\ 
   &  & LASSO & 15.00(0.00) & 76.10(27.42) & 0.30(0.07) & 0.41(0.06) & 3.49(0.78) \\ 
				\cline{2-8}
				& Error2 & ssQLASSO & 14.73(0.58) & 0.70(1.12) & 0.97(0.04) & 0.97(0.04) & 4.77(1.05) \\ 
   &  & QLASSO & 15.00(0.00) & 49.40(5.10) & 0.38(0.02) & 0.48(0.02) & 4.71(0.48) \\ 
   &  & ssLASSO & 8.80(3.12) & 0.23(0.50) & 0.71(0.16) & 0.74(0.13) & 5.54(1.87) \\ 
   &  & LASSO & 13.67(3.13) & 73.27(38.46) & 0.28(0.09) & 0.38(0.10) & 9.97(2.52) \\ 
				\cline{2-8}
				& Error3 & ssQLASSO & 14.90(0.31) & 0.17(0.38) & 0.99(0.01) & 0.99(0.01) & 4.06(0.88) \\ 
   &  & QLASSO & 15.00(0.00) & 50.97(6.81) & 0.37(0.03) & 0.47(0.03) & 4.07(0.60) \\ 
   &  & ssLASSO & 9.33(3.46) & 0.03(0.18) & 0.74(0.19) & 0.77(0.16) & 4.75(2.17) \\ 
   &  & LASSO & 15.00(0.00) & 84.90(30.72) & 0.28(0.07) & 0.39(0.06) & 8.01(2.07) \\
				\cline{2-8}
				& Error4 & ssQLASSO & 14.43(0.90) & 1.27(1.53) & 0.94(0.05) & 0.94(0.04) & 5.02(1.03) \\ 
   &  & QLASSO & 15.00(0.00) & 49.50(5.39) & 0.38(0.03) & 0.48(0.02) & 5.55(0.79) \\ 
   &  & ssLASSO & 8.50(3.70) & 0.23(0.50) & 0.68(0.21) & 0.72(0.17) & 5.54(2.29) \\ 
   &  & LASSO & 15.00(0.00) & 71.60(25.68) & 0.31(0.07) & 0.42(0.06) & 7.86(1.52) \\
				\cline{2-8}
				& Error5 & ssQLASSO & 14.57(1.04) & 0.90(1.67) & 0.96(0.06) & 0.96(0.05) & 4.07(0.92) \\ 
   &  & QLASSO & 15.00(0.00) & 51.13(6.43) & 0.37(0.03) & 0.47(0.02) & 5.74(0.76) \\ 
   &  & ssLASSO & 10.20(3.84) & 0.67(0.96) & 0.76(0.17) & 0.79(0.15) & 4.39(2.22) \\ 
   &  & LASSO & 15.00(0.00) & 81.53(33.82) & 0.29(0.07) & 0.40(0.06) & 7.16(2.07) \\
				\hline
				$\tau=0.7$& Error1 & ssQLASSO & 15.00(0.00) & 1.27(1.70) & 0.96(0.05) & 0.96(0.04) & 0.90(0.19) \\ 
   &  & QLASSO & 15.00(0.00) & 36.43(7.11) & 0.46(0.05) & 0.54(0.04) & 3.08(0.31) \\ 
   &  & ssLASSO & 15.00(0.00) & 0.67(1.32) & 0.98(0.04) & 0.98(0.04) & 0.66(0.16) \\ 
   &  & LASSO & 15.00(0.00) & 67.73(22.53) & 0.32(0.08) & 0.43(0.07) & 3.21(0.52) \\ 
				\cline{2-8}
				& Error2 & ssQLASSO & 14.63(0.61) & 0.73(1.01) & 0.96(0.03) & 0.96(0.03) & 3.34(1.29) \\ 
   &  & QLASSO & 14.97(0.18) & 33.73(3.90) & 0.47(0.03) & 0.55(0.02) & 5.41(0.73) \\ 
   &  & ssLASSO & 8.70(3.60) & 0.13(0.35) & 0.70(0.21) & 0.74(0.17) & 5.34(2.29) \\ 
   &  & LASSO & 14.33(2.59) & 64.37(24.21) & 0.31(0.08) & 0.42(0.07) & 8.23(1.71) \\ 
				\cline{2-8}
				& Error3 & ssQLASSO & 14.93(0.25) & 0.30(0.47) & 0.99(0.02) & 0.99(0.02) & 2.87(0.90) \\ 
   &  & QLASSO & 14.97(0.18) & 33.40(5.30) & 0.48(0.04) & 0.55(0.03) & 5.63(0.72) \\ 
   &  & ssLASSO & 12.33(2.37) & 0.30(0.60) & 0.88(0.10) & 0.89(0.09) & 3.11(1.41) \\ 
   &  & LASSO & 14.87(0.57) & 70.40(26.08) & 0.31(0.07) & 0.42(0.06) & 7.31(1.82) \\ 
				\cline{2-8}
				& Error4 & ssQLASSO & 14.10(1.21) & 1.10(2.43) & 0.94(0.07) & 0.94(0.06) & 3.86(1.64) \\ 
   &  & QLASSO & 14.93(0.25) & 34.20(5.16) & 0.47(0.03) & 0.55(0.03) & 6.54(1.07) \\ 
   &  & ssLASSO & 7.20(3.18) & 0.17(0.38) & 0.62(0.19) & 0.67(0.16) & 6.17(1.81) \\ 
   &  & LASSO & 14.93(0.25) & 74.17(32.31) & 0.31(0.07) & 0.41(0.06) & 8.13(2.41) \\
				\cline{2-8}
				& Error5 & ssQLASSO & 14.80(0.41) & 1.03(1.47) & 0.96(0.05) & 0.96(0.05) & 3.53(1.10) \\ 
   &  & QLASSO & 14.80(0.48) & 33.90(5.18) & 0.47(0.04) & 0.54(0.03) & 6.66(1.12) \\ 
   &  & ssLASSO & 9.53(4.11) & 0.87(3.10) & 0.72(0.24) & 0.77(0.16) & 4.76(2.44) \\ 
   &  & LASSO & 15.00(0.00) & 73.70(28.40) & 0.31(0.07) & 0.41(0.06) & 6.87(1.77) \\
				\hline
			\end{tabular}
		}
	\end{center}
	\centering
\end{table}

We have performed additional simulations beyond the aforementioned 8 case scenarios. First, we consider generating nonzero coefficients that also include negative ones. In particular, we use $U[0.6,0.8]$ and $U[-0.8,-0.6]$ to generate 8 and 7 nonzero coefficients in coefficient vector $\boldsymbol \beta$, respectively. Under homogeneous model error, AR-1 correlation and $p$=1600, Table \ref{sim9} demonstrates similar patterns in favor of ssQLASSO. Besides, we have investigated the performance when the correlation structure has been extracted from the real genetic data. At each replicate, we randomly draw 1600 gene expressions from the TCGA SKCM data analyzed in the case study. The correlation of those genes are adopted to generate predictors in the homogeneous model. In contrast to Table \ref{sim2} and \ref{sim4} that are under AR1 and banded correlation correspondingly, we can observe a clear drop in performance of all methods under comparison in Table \ref{sim10}. Interestingly, ssQLASSO has even outperformed ssLASSO under $i.i.d.$ $N$(0,1) error in terms of better F1 score, MCC and estimation error. For example, with 50$\%$ quantile level, those measures for the proposed ssQLASSO are 0.92 (sd0.06), 0.92 (sd0.05) and 1.58 (sd0.49), respectively. While ssLASSO has yielded an F1 score of 0.45 (sd0.14), an MCC of 0.51 (sd0.12) and L1 error of 8.82 (sd1.51). The ssLASSO has significantly deteriorated even under $N$(0,1) error. Furthermore, we have conducted sensitivity analysis on ssQLASSO when using different hyper-parameters $(a,b)$ to specify the inverse-gamma prior on $\sigma$. With $p$=1600 and the AR-1 correlation, we have assessed the performance using $(a,b)=(1,5)$, $(0.5,3)$ and $(5,1)$ for quantile level $\tau$=0.3, 0.5 and 0.7, respectively, in both Table \ref{sim11} under $i.i.d.$ error and Table \ref{sim12} under non-$i.i.d.$ error. There is no deviations from the patterns shown in the counterpart Table \ref{sim2} and \ref{sim6} under $(a,b)=(1,1)$. Therefore, the performance of ssQLASSO is not sensitive to different choices of hyper-parameters.

\begin{figure}[h]
	\begin{center}
		\includegraphics[height=0.3\textheight]{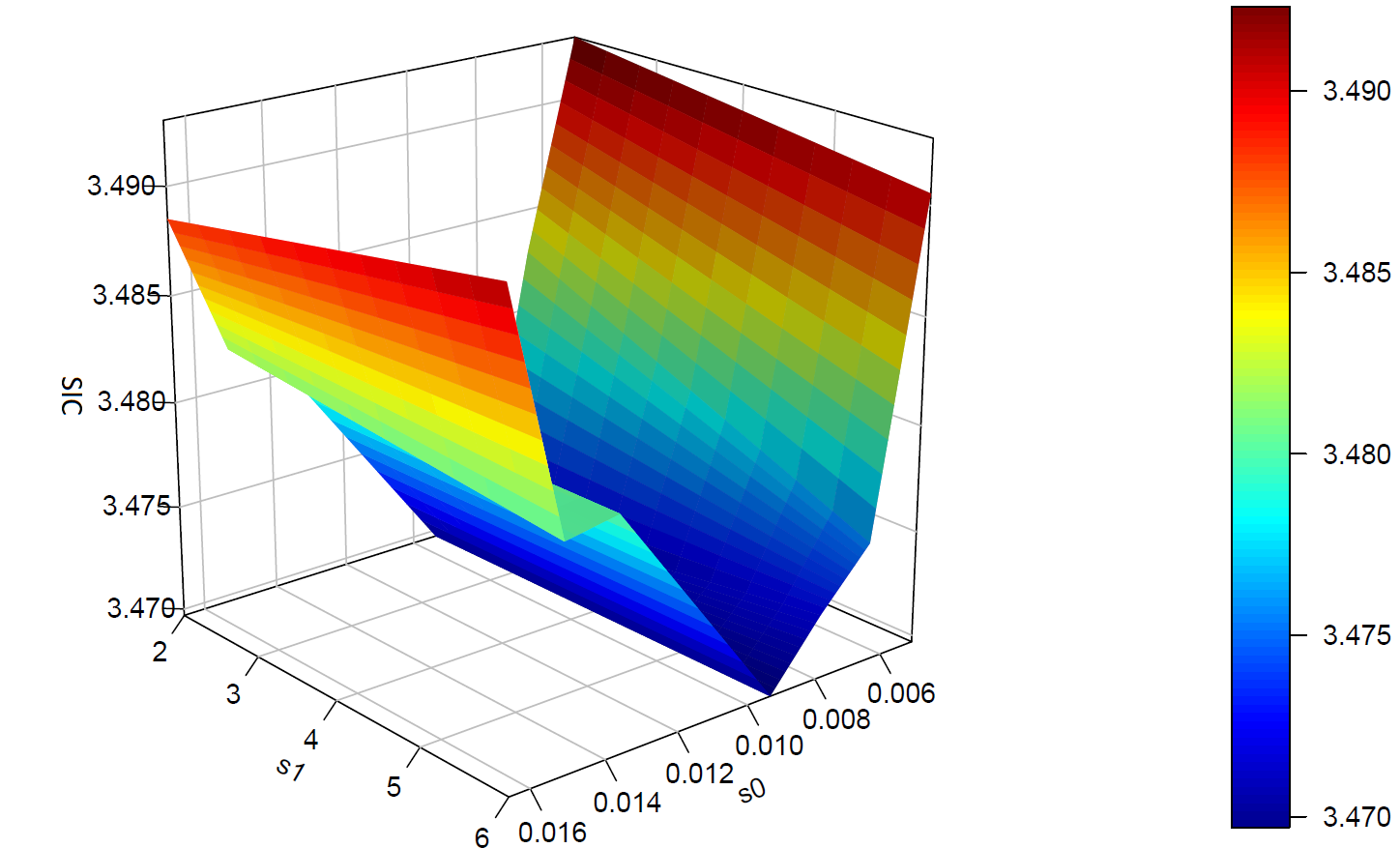}
		\caption{\footnotesize The surface plot of Schwarz Information Criterion (SIC) using ssQLASSO under $\tau=30\%$, AR-1 correlation, $(n=400,p=1600)$ and error 3 over 100 replications.}\label{SIC}
	\end{center}
\end{figure}

\textbf{Tuning parameter selection}: The performance of ssQLASSO critically depends on the tuning parameters $s_0$ and $s_1$, the scales for the spike and slab components, correspondingly. In our simulation study, the optimal pair of ($s_0,s_1$) has been chosen via the Schwarz Information
Criterion (SIC). Figure \ref{SIC} has revealed the V-shape ``valley" of the 3D surface plot of SIC. Compared to the spike scale $s_0$ that determines the sparsity of ssQLASSO, we can see that the choices of $s_1$ is relatively less important. As the slab scale $s_1$ changes from 4 to 6, the smallest SIC criterion value on the V-shape ``valley" do not have too much variations if the spike scale $s_0$ has been selected appropriately. Such a pattern is similar to the surface plot of prediction error of ssLASSO in published studies \cite{tang2017spike,tang2017spikeGLM}, and explains why the strategy of fixing the slab scale $s_1$ (e.g. $s_1$=2) and conducting a fine tune of $s_0$ over a grid of values has been adopted. We have found out that there is not much difference between such a strategy and our search of optimal pair of tunings. In the Appendix, we have provided the 3D surface plot of cross-validation error in terms of quantile check loss in Figure \ref{3d}, which shows the V-shape ``valley" similar to Figure \ref{SIC}. Consequently, the superior performance of ssQLASSO observed in the simulation study is insensitive to the strategies adopted to locate optimal tuning parameters.

\textbf{Solution path and self-adaptivity}: The impact of the spike scale $s_0$ on sparsity can be thoroughly understood by examining the solution path of ssQLASSO. With ($n$,$p$)=(400,1600) and AR1 correlation, we generate the response from homogeneous model under Error3 and quantile level $\tau=0.5$, by setting 5 coefficients in vector $\boldsymbol \beta$ as nonzero, i.e., (0.8, 0.6, 0.4, -0.5, -0.6). When fitting the data using ssQLASSO, we fix the slab scale $s_1$ to 1. The solution paths corresponding to LASSO and ssQLASSO are shown in Figure \ref{LASSOSP} and \ref{BQLSSSP} in Appendix, respectively. For illustration purpose, we only present paths corresponding to the aforementioned 5 nonzero coefficients. As $\lambda$ increases, all the coefficient profiles of LASSO estimates in Figure \ref{LASSOSP} converge to the horizontal axis at 0, regardless of the nonzero coefficients or not. On the contrary,  Figure \ref{BQLSSSP} shows that with $s_0$ growing beyond the optimal tuning around 0.01, those important coefficients are consistently included in the model with little or no shrinkage. 

Therefore, unlike LASSO, the ssQLASSO is self-adaptive to underlying sparsity pattern even when an excessive amount of penalty is imposed. The slab component is essential in preventing coefficients with large magnitude from being absorbed by the spike component, so the coefficient trajectories still stabilize as the amount of shrinkage increases. Otherwise the pattern of solution path will be the same as that of Figure \ref{LASSOSP}. The self-adaptivity also distinguishes the selective shrinkage mechanism of ssQLASSO from that of adaptive (quantile) LASSO which shrinks all the coefficients to 0 under large penalty.

\textbf{Computational time}: The computational advantage of ssQLASSO over the alternatives can be better understood by assessing the run time using the setting of Table \ref{sim2}, that is, a homogeneous data generating model with AR-1 correlation and $(n=400, p=1600)$. For a fair comparison, the run time in seconds is recorded under fixed tuning parameters, as shown in Table \ref{rt} from the Appendix. Overall, the computational cost of ssQLASSO is much less than QLASSO and comparable to its non-robust counterpart ssLASSO. For example, at $\tau$=0.3 and the mixture normal error (Error4), the run time of ssQLASSO is 24.19(sd3.84), which is comparable to 10.43(sd0.56) of ssLASSO and much smaller than 264.00(sd10.66) of QLASSO. 

Robust regularization with non-differentiable loss function incurs more cost in computation compared to non-robust methods with smooth losses, as shown by the run time difference between QLASSO and LASSO in Table \ref{rt}. However, the run time of ssQLASSO and ssLASSO are of the same order, much less than that of the quantile LASSO (QLASSO). Such a phenomenon is due to formulating quantile regression with the asymmetric Laplace distribution which results in the quadratic form of the objective function in equation (\ref{equr:lBQLSS}). Therefore, the estimation of high-dimensional coefficient vector $\boldsymbol \beta$ can be conducted through the fast coordinate descent update of the soft-thresholding rule \cite{friedman2010regularization} which is not attainable under non-differentiable loss.

\textbf{Additional simulations}: We have further assessed the performance of ssQLASSO and alternatives through additional simulations in Section F from the Appendix. In Section F.1, we have considered much larger settings  $(n,p)=(800,2400)$ and (800,3200) in Tables \ref{homo2400} -- \ref{heter3200}, as well as the settings using data generating models incorporating the clinical covariates not under selection in Table \ref{homoclinic} and Table \ref{heterclinc}. The advantage of ssQLASSO over alternatives still stands. Accordingly, the computational times of the proposed method under growing study sizes have been shown in Figure \ref{Ctime}. In Section F.3, we have evaluated the appropriateness of adopting the asymmetric Laplace distribution (ALD) for the likelihood function. The maximum overlapping density plots in Figure \ref{density1} and Figure \ref{density2} justify the utility of ALD when distributional assumptions are violated. The summary statistics on number of outliers in simulation are provided in Table \ref{outliers} in Section F.5, indicating adaptivity of ALD under a larger numbers of outliers generated from the heterogeneous models. Overall, the additional simulation studies have further demonstrated the superiority of ssQLASSO over competing methods.

\section{Real Data Analysis}

The Cancer Genome Atlas (TCGA, \url{https://cancergenome.nih.gov/}) is a joint effort from the National Cancer Institute and the National Human Genome Research Institute to provide high-quality molecular profiling data for 33 cancer types. We have analyzed the Cancer Genome Atlas (TCGA) lung adenocarcinomas (LUAD) and skin cutaneous melanoma (SKCM) data using level 3 gene expression measurements
downloaded from the cBio Cancer Genomics Portal (Cerami et al., 2012)\cite{cerami2012cbio}.

\subsection{TCGA lung adenocarcinomas (LUAD) data}
Lung cancer is the leading cause of cancer-related deaths in the United States. LUAD is one of the most common subtype of lung cancer, accounting for  approximately 40$\%$ of all lung cancer cases. With the TCGA LUAD data (Campbell et al. (2011)\cite{campbell2016distinct}), we treat the variable ``forced expiratory volume in 1 second", or FEV1, as the disease phenotype, with the corresponding histogram shown in Figure \ref{hists}. Age, gender, smoking history and pack per year have been included as clinical covariates, where gender and smoking history are categorical, and the other two are continuous. We match the gene expression measurements with the phenotypic trait and clinical covariates through common subjects, resulting in a working dataset with a total of 208 subjects and 20,531 gene expressions. Following published studies (\cite{li2015bayesian,wu2014integrative}), we first conduct pre-screening to reduce the computational cost. The gene expressions have been ranked through coefficient of variations, and we have selected top 1200 genes for downstream analysis.

We have set quantile level $\tau=50\%$ for ssQLASSO and QLASSO, and applied all four methods to the LUAD dataset. Table \ref{real} shows the pairwise overlap of findings, demonstrating the difference in identification results between the proposed method and alternatives. Specifically, the proposed ssQLASSO has identified 17 genes which are LCOR, ALPL, AATBC, AADAT, AGPAT3, POLR3G, C6orf62, UCK2, NUDT16L1, SMC3, SRP14, TMEM217, RPP25 and DDO. In particular, there are only 3 genes commonly discovered by ssQLASSO and SSLASSO, indicating that the spike-and-slab quantile LASSO prior leads to a quite different set of findings when heterogeneity in disease trait exists. Among these genes that have been selected by ssQLASSO, SMC3, SRP14 and RPP25 have not been reported by other approaches. Wang et al., 2018 \cite{wang2018hydrogen} have confirmed that the progression of lung cancer can be inhibited via targeting SMC3, or the structural maintenance of chromosomes 3, through hydrogen gas. Malik et al., 2019 \cite{malik2019micrornas} have reported  SRP14 as a potential biomarker of LUAD, which is consistent with its changed expression levels in cell lines with KRAS mutations. In addition, Mecoli et al., 2021 \cite{mecoli2021cancer} have shown that patients with antibodies against RPP25 are significantly less likely to develop cancer within 2 years of systemic sclerosis (SSc) onset which are frequently caused by lung disease (Steen and Medsger, 2007)\cite{steen2007changes}. 

We have evaluated the prediction accuracy based on multi-split of the data. At
each split, three-fourth of the subjects are randomly chosen as the training set for model fitting, while the remaining data is utilized as the testing set to assess prediction performance. Then the mean and standard deviation of prediction errors can be computed over the 100 splits. In practice, prediction performance of robust and non-robust methods are usually evaluated using different criteria, such as the prediction mean squared errors (PMSE) for ssLASSO and LASSO, and prediction mean absolute deviations (PMAD) for ssQLASSO and QLASSO at the $50\%$ quantile level. For a fair cross-comparison, we have computed both PMAD and PMSE for all the methods. The PMAD of the robust ssQLASSO and QLASSO are 13.51 (sd 1.84) and 15.61 (sd 1.69), respectively. Their corresponding PMSEs are 327.38 (sd 90.35) and 435.86 (sd 94.78). For non-robust methods ssLASSO and LASSO, the PMADs are 15.65 (sd 2.02) and 17.65 (sd 1.91), and the PMSEs are 419.52 (sd 113.39) and 557.56 (sd 125.24), respectively. The proposed ssQLASSO has the smallest prediction error in terms of both PMADs and PMSEs. The first row of Figure \ref{PP} listed the box-plots of PMADs and PMSEs of four methods across 100 splits, which clearly indicate the superior performance of ssQLASSO in terms of prediction. Particularly, it appears that ssQLASSO is more stable as there are no outlying prediction errors as shown in other box-plots. We can obtain the same conclusions regarding the better performance of ssQLASSO in Section F.5 from the Appendix when analyses at quantile levels 0.3 and 0.7 have been conducted with overlapping results shown in Table \ref{real1}. 
 
\subsection{TCGA skin cutaneous melanoma (SKCM) data}

Cutaneous melanoma (CM) is the most prevalent form of melanoma across the world. Its incidence has risen steadily, especially among young adults. While CM only comprises of 3$\%$ of all skin cancer cases, it causes 65$\%$ of skin cancer deaths \cite{naik2021cutaneous}. In CM, Breslow's thickness is one of the most important prognostic factors indicating the progression of the disease. Figure \ref{hists} shows that in TCGA SKCM, the Breslow’s depth is skewed and has outlying observations even on the log scale. In the case study, we use the log-transformed Breslow’s depth as the response variable. The clinical factors are age, initial cancer year and gender, where the first two are continuous and gender is categorical. The initial dataset after matching the phenotype with the genotypes consists of 353 subjects and 20,531 gene expressions. Similar to the LUAD study, we have selected the top 1200 genes for further analysis.

For ssQLASSO and QLASSO, we set the quantile level $\tau$ to $50\%$ and analyze the SKCM dataset using all methods under comparison. The number of identified and overlapping genes are shown in the right panel of Table \ref{real}, revealing  smaller numbers of overlapping genes in general, compared to those from analysis of LUAD. In particular, ssQLASSO and QLASSO have only commonly identified 2 genes, which indicates the challenge of analysis in the presence of heavily skewed phenotypic trait and  the necessity of developing robust variable selection methods. The proposed ssQLASSO methods have identified 13 important genes: TTC28-AS1, NPR1, CEMIP, FLJ20373, LOC126807, GJA3, INPP1, FMNL2, COQ9, ZFAT, KCTD7, KCTD21 and TFR2. Among these genes, MAP4K4, GJA3, INPP1 and KCTD7 have not been found by the alternatives. Surman et al., 2023\cite{surman2023similarities} have reported that during the metastasis of cutaneous melanoma (CM), the CM-derived exosomes may have contributions to organ-specific metastasis, and the MAP4K4 are in presence of majority of the CM-derived exosome samples. Orellana et al.2021\cite{orellana2021connexins}  have found that melanoma patients with higher GJA3 tumor expression levels have a much worse prognosis than those exhibiting lower GJA3 tumor expressions. As for INPP1, Benjamin et al., 2014\cite{benjamin2014inositol} have reported the up-regulation of INPP1 in melanoma tumors compared to normal skin tissues. Moreover, Angrisani et al., 2021\cite{angrisani2021emerging} have shown that KCTD7 is up-regulated in tumors influencing skin tissues. 

\begin{figure}[H]
		\begin{center}
		\includegraphics[height=0.3\textheight]{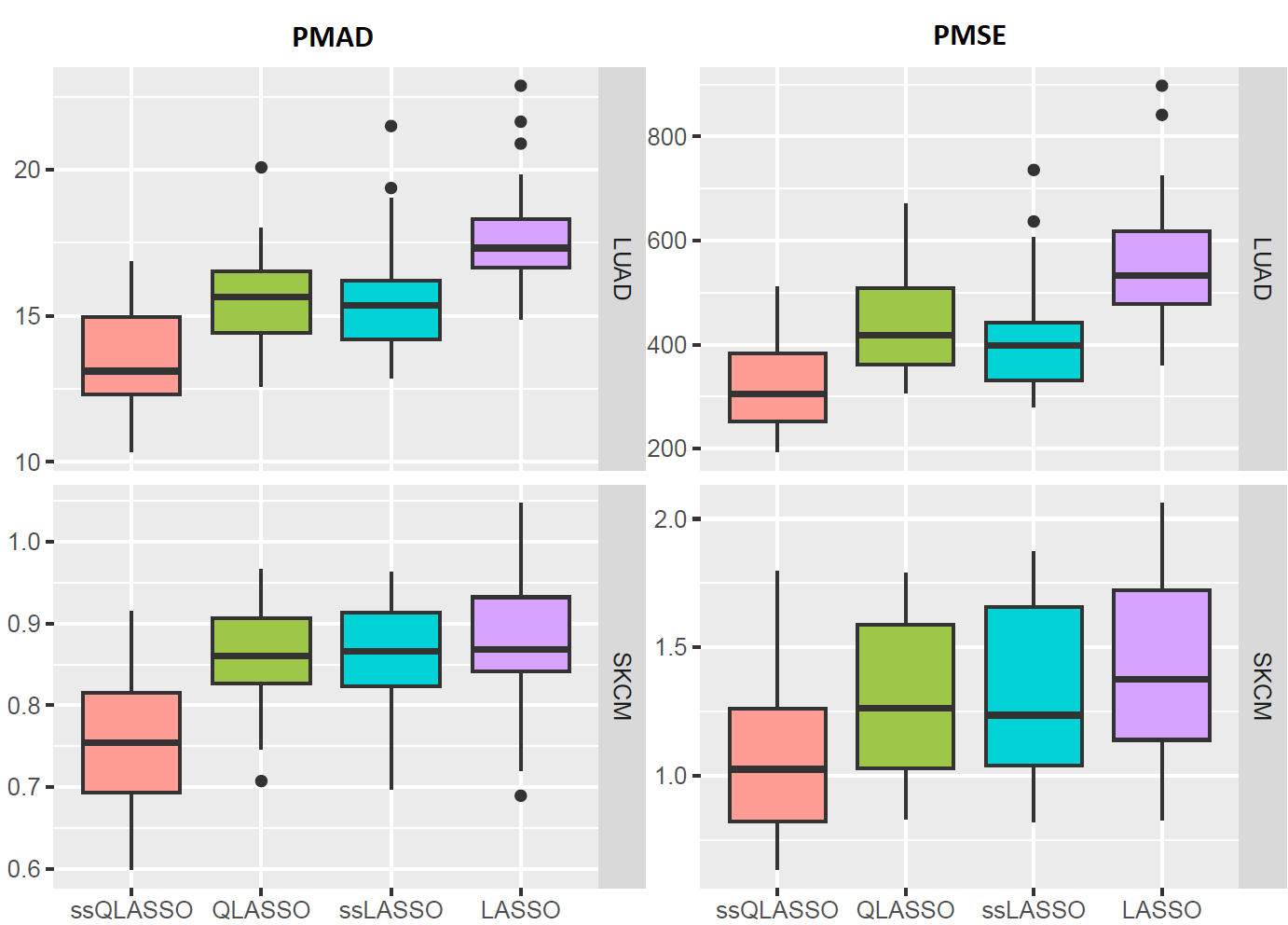}
		\caption{\footnotesize PMAD (the first column) and PMSE (the second column) corresponding to the LUAD (the first row) and SKCM (the second row).}\label{PP}
		\end{center}
	\end{figure}

Since it is difficult to have an objective assessment of the selection accuracy in real data, we conduct multiple data splitting to partially justify the utility of ssQLASSO. For robust methods ssQLASSO and QLASSO, the PMADs are 0.76 (sd 0.08) and 0.86 (sd 0.06) and the PMSEs are 1.07 (sd 0.33) and 1.41 (sd 0.33), respectively. For non-robust methods ssLASSO and LASSO, the PMADs are 0.86 (sd 0.07) and 0.88 (sd 0.08) and the PMSEs are 1.41 (sd 0.34) and 1.42 (sd 0.36), respectively.  The box-plots of PMSE and PMAD across 100 splits for SKCM data are shown in the second row of Figure \ref{PP}, which has demonstrated the better prediction performance of ssQLASSO over alternatives in both criteria. Consistent observations on the improved performance of ssQLASSO can be observed in Section F.5 from the Appendix under quantile levels 0.3 and 0.7. The overlapping results are provided in Table \ref{real2}. 


\begin{table} [H]
	\def\arraystretch{1.5}
\begin{center}
\fontsize{9}{9}\selectfont{
    \caption{Identification results for real data analysis. The numbers of genetic effects identified by different approaches and their overlaps.}\label{real}
    \centering
\begin{tabular}{ l l l l l l l l l l}
&LUAD&&&&&SKCM&&&
\\
\cline{2-5}\cline{7-10}
&ssQLASSO&QLASSO&ssLASSO&LASSO&&ssQLASSO&QLASSO&ssLASSO&LASSO
\\
ssQLASSO&17 &12 &3 &7 & &13 &2	&4&8
\\
QLASSO&&30 &6 &9 & & &17 &1&6
\\
ssLASSO&&&12 &9 & & & &8&4
\\
LASSO&&&&51 & & & &&27
\\
\hline
\end{tabular}
}
\end{center}
\centering
\end{table}

\FloatBarrier

\section{Discussion}

The main appeal of the ssQLASSO lies in quickly finding the posterior mode in robust Bayesian analysis, leading to a much smaller computational cost compared to MCMCs. In literature, the variational Bayes methods and related ones also suffice such an analysis goal \cite{carbonetto2012scalable,komodromos2022variational,wang2020simple}. A direct comparison between the two types of methods worths further explorations, particularly under disease traits with skewed distributions and heavy-tailed errors. Recently, it has been revealed that the spike--and--slab LASSO can be developed for approximate posterior sampling and uncertainty quantification based on Bayesian bootstrap \cite{nie2023bayesian}, which is potentially applicable for extending ssQLASSO to approximate the posterior distributions beyond the mode detection.          
 
Analysis of convergence rate of EM algorithms in high-dimensional settings remains an open problem with sparse literature available. Although such a rate has not been shown for spike--and--slab LASSO yet, we speculate that ssQLASSO will demonstrate a similar convergence pattern since update of the high-dimensional coefficient vector $\boldsymbol \beta$ in the M step can be done using the same coordinate-descent strategy as in ssLASSO. To handle ultra-high-dimensional predictors such as in the GWAS, we suggest a marginal screening of the predictors to a reasonable scale so ssQLASSO and alternative methods can be applied, which has been widely acknowledged and adopted in ultra-high dimensional studies \cite{fan2008sure,wang2012quantile}. 

The robustness of ssQLASSO comes from its resistance to outliers and data contamination in the response variable, such as disease phenotypes. Since it assumes linear structure on predictors, the proposed method may lead to inferior performances when the model is misspecified due to structured sparsity such as interaction or group structures \cite{huang2012selective,zhou2021gene}. We conjecture that it can be further extended to handle more complicated underlying data structures. For example, a group spike-and-slab quantile LASSO can be developed to incorporate group structures. 

Last but not least, quantile regression has been historically proposed to dissect heterogeneous data through estimation at a fixed quantile level \cite{koenker1978regression}. It is a well acknowledged limitation of quantile regression, especially under high-dimensional scenarios, to achieve a final set of predictors across different quantile levels \cite{wu2009variable,wang2012quantile,li2010bayesian,wang2009quantile}. We follow the common practice of published studies to report results at different values of $\tau$ when a specific quantile level of interest is lacking. Such an analysis can be improved by considering composite quantile regression and modeling multiple quantile levels simultaneously \cite{zou2008composite}. Specifically, we conjecture that the Bayesian composite quantile regression can be incorporated in the proposed ssQLASSO \cite{huang2015bayesian}.


\section{Conflict of Interest}
The authors declare that there is no conflict of interest.

\section{Acknowledgements}
This work was partially supported by an Innovative Research Award from the Johnson Cancer Research Center at Kansas State University and the National Institutes of Health (NIH) grant R01 CA204120.

\bibliography{references}

\begin{thebibliography}{10}

\bibitem{rovckova2018spike}
V.~Ro{\v{c}}kov{\'a} and E.~I. George, ``The spike-and-slab lasso,'' {\em
  Journal of the American Statistical Association}, vol.~113, no.~521,
  pp.~431--444, 2018.

\bibitem{cerami2012cbio}
E.~Cerami, J.~Gao, U.~Dogrusoz, B.~E. Gross, S.~O. Sumer, B.~A. Aksoy,
  A.~Jacobsen, C.~J. Byrne, M.~L. Heuer, E.~Larsson, {\em et~al.}, ``The cbio
  cancer genomics portal: an open platform for exploring multidimensional
  cancer genomics data,'' {\em Cancer Discovery}, vol.~2, no.~5, pp.~401--404,
  2012.

\bibitem{campbell2016distinct}
J.~D. Campbell, A.~Alexandrov, J.~Kim, J.~Wala, A.~H. Berger, C.~S. Pedamallu,
  S.~A. Shukla, G.~Guo, A.~N. Brooks, B.~A. Murray, {\em et~al.}, ``Distinct
  patterns of somatic genome alterations in lung adenocarcinomas and squamous
  cell carcinomas,'' {\em Nature Genetics}, vol.~48, no.~6, pp.~607--616, 2016.

\bibitem{marghoob2000breslow}
A.~A. Marghoob, K.~Koenig, F.~V. Bittencourt, A.~W. Kopf, and R.~S. Bart,
  ``Breslow thickness and clark level in melanoma: support for including level
  in pathology reports and in american joint committee on cancer staging,''
  {\em Cancer}, vol.~88, no.~3, pp.~589--595, 2000.

\bibitem{wu2015selective}
C.~Wu and S.~Ma, ``A selective review of robust variable selection with
  applications in bioinformatics,'' {\em Briefings in Bioinformatics}, vol.~16,
  no.~5, pp.~873--883, 2015.

\bibitem{alfons2013sparse}
A.~Alfons, C.~Croux, and S.~Gelper, ``Sparse least trimmed squares regression
  for analyzing high-dimensional large data sets,'' {\em The Annals of Applied
  Statistics}, pp.~226--248, 2013.

\bibitem{ren2019robust}
J.~Ren, Y.~Du, S.~Li, S.~Ma, Y.~Jiang, and C.~Wu, ``Robust network-based
  regularization and variable selection for high-dimensional genomic data in
  cancer prognosis,'' {\em Genetic Epidemiology}, vol.~43, no.~3, pp.~276--291,
  2019.

\bibitem{bradic2011penalized}
J.~Bradic, J.~Fan, and W.~Wang, ``Penalized composite quasi-likelihood for
  ultrahigh dimensional variable selection,'' {\em Journal of the Royal
  Statistical Society Series B: Statistical Methodology}, vol.~73, no.~3,
  pp.~325--349, 2011.

\bibitem{daye2012high}
Z.~J. Daye, J.~Chen, and H.~Li, ``High-dimensional heteroscedastic regression
  with an application to eqtl data analysis,'' {\em Biometrics}, vol.~68,
  no.~1, pp.~316--326, 2012.

\bibitem{peng2015iterative}
B.~Peng and L.~Wang, ``An iterative coordinate descent algorithm for
  high-dimensional nonconvex penalized quantile regression,'' {\em Journal of
  Computational and Graphical Statistics}, vol.~24, no.~3, pp.~676--694, 2015.

\bibitem{gao2010robust}
X.~Gao and J.~Huang, ``A robust penalized method for the analysis of noisy dna
  copy number data,'' {\em BMC genomics}, vol.~11, no.~1, pp.~1--10, 2010.

\bibitem{wang2011identification}
H.~J. Wang and J.~Hu, ``Identification of differential aberrations in
  multiple-sample array cgh studies,'' {\em Biometrics}, vol.~67, no.~2,
  pp.~353--362, 2011.

\bibitem{wu2018robust}
C.~Wu, Q.~Zhang, Y.~Jiang, and S.~Ma, ``Robust network-based analysis of the
  associations between (epi) genetic measurements,'' {\em Journal of
  Multivariate Analysis}, vol.~168, pp.~119--130, 2018.

\bibitem{cohen2019robust}
G.~V. Cohen~Freue, D.~Kepplinger, M.~Salibi{\'a}n-Barrera, and E.~Smucler,
  ``Robust elastic net estimators for variable selection and identification of
  proteomic biomarkers,'' {\em Annals of Applied Statistics}, vol.~13, no.~4,
  pp.~2065--2090, 2019.

\bibitem{kepplinger2021robust}
D.~Kepplinger and G.~V. Cohen~Freue, ``Robust prediction and protein selection
  with adaptive pense,'' in {\em Statistical Analysis of Proteomic Data:
  Methods and Tools}, pp.~315--331, Springer, 2021.

\bibitem{o2009review}
R.~B. O'hara and M.~J. Sillanp{\"a}{\"a}, ``A review of bayesian variable
  selection methods: what, how and which,'' {\em Bayesian Analysis}, vol.~4,
  no.~1, pp.~85--117, 2009.

\bibitem{ickstadt2018toward}
K.~Ickstadt, M.~Sch{\"a}fer, and M.~Zucknick, ``Toward integrative bayesian
  analysis in molecular biology,'' {\em Annual Review of Statistics and Its
  Application}, vol.~5, pp.~141--167, 2018.

\bibitem{johndrow2020scalable}
J.~Johndrow, P.~Orenstein, and A.~Bhattacharya, ``Scalable approximate mcmc
  algorithms for the horseshoe prior,'' {\em Journal of Machine Learning
  Research}, vol.~21, no.~73, pp.~1--61, 2020.

\bibitem{liang2022adaptive}
X.~Liang, S.~Livingstone, and J.~Griffin, ``Adaptive random neighbourhood
  informed markov chain monte carlo for high-dimensional bayesian variable
  selection,'' {\em Statistics and Computing}, vol.~32, no.~5, p.~84, 2022.

\bibitem{liang2023adaptive}
X.~Liang, S.~Livingstone, and J.~Griffin, ``Adaptive mcmc for bayesian variable
  selection in generalised linear models and survival models,'' {\em Entropy},
  vol.~25, no.~9, p.~1310, 2023.

\bibitem{lu2021identifying}
X.~Lu, K.~Fan, J.~Ren, and C.~Wu, ``Identifying gene--environment interactions
  with robust marginal bayesian variable selection,'' {\em Frontiers in
  Genetics}, vol.~12, p.~667074, 2021.

\bibitem{ren2023robust}
J.~Ren, F.~Zhou, X.~Li, S.~Ma, Y.~Jiang, and C.~Wu, ``Robust bayesian variable
  selection for gene--environment interactions,'' {\em Biometrics}, vol.~79,
  no.~2, pp.~684--694, 2023.

\bibitem{zhou2023bayesian}
F.~Zhou, J.~Ren, S.~Ma, and C.~Wu, ``The bayesian regularized quantile varying
  coefficient model,'' {\em Computational Statistics \& Data Analysis},
  p.~107808, 2023.

\bibitem{im2023bayesian}
Y.~Im, Y.~Huang, A.~Tan, and S.~Ma, ``Bayesian finite mixture of regression
  analysis for cancer based on histopathological imaging--environment
  interactions,'' {\em Biostatistics}, vol.~24, no.~2, pp.~425--442, 2023.

\bibitem{yu2001bayesian}
K.~Yu and R.~A. Moyeed, ``Bayesian quantile regression,'' {\em Statistics \&
  Probability Letters}, vol.~54, no.~4, pp.~437--447, 2001.

\bibitem{li2010bayesian}
Q.~Li, N.~Lin, and R.~Xi, ``Bayesian regularized quantile regression,'' {\em
  Bayesian Analysis}, vol.~5, no.~3, pp.~533--556, 2010.

\bibitem{george1993variable}
E.~I. George and R.~E. McCulloch, ``Variable selection via gibbs sampling,''
  {\em Journal of the American Statistical Association}, vol.~88, no.~423,
  pp.~881--889, 1993.

\bibitem{brown2010inference}
P.~J. Brown and J.~E. Griffin, ``Inference with normal-gamma prior
  distributions in regression problems,'' {\em Bayesian Analysis}, vol.~5,
  no.~1, pp.~171--188, 2010.

\bibitem{carvalho2009handling}
C.~M. Carvalho, N.~G. Polson, and J.~G. Scott, ``Handling sparsity via the
  horseshoe,'' in {\em Artificial Intelligence and Statistics}, pp.~73--80,
  PMLR, 2009.

\bibitem{friedman2010regularization}
J.~Friedman, T.~Hastie, and R.~Tibshirani, ``Regularization paths for
  generalized linear models via coordinate descent,'' {\em Journal of
  Statistical Software}, vol.~33, no.~1, p.~1, 2010.

\bibitem{tang2017spike}
Z.~Tang, Y.~Shen, X.~Zhang, and N.~Yi, ``The spike-and-slab lasso cox model for
  survival prediction and associated genes detection,'' {\em Bioinformatics},
  vol.~33, no.~18, pp.~2799--2807, 2017.

\bibitem{tang2017spikeGLM}
Z.~Tang, Y.~Shen, X.~Zhang, and N.~Yi, ``The spike-and-slab lasso generalized
  linear models for prediction and associated genes detection,'' {\em
  Genetics}, vol.~205, no.~1, pp.~77--88, 2017.

\bibitem{kozumi2011gibbs}
H.~Kozumi and G.~Kobayashi, ``Gibbs sampling methods for bayesian quantile
  regression,'' {\em Journal of Statistical Computation and Simulation},
  vol.~81, no.~11, pp.~1565--1578, 2011.

\bibitem{wang2007robust}
H.~Wang, G.~Li, and G.~Jiang, ``Robust regression shrinkage and consistent
  variable selection through the lad-lasso,'' {\em Journal of Business \&
  Economic Statistics}, vol.~25, no.~3, pp.~347--355, 2007.

\bibitem{park2008bayesian}
T.~Park and G.~Casella, ``The bayesian lasso,'' {\em Journal of the American
  Statistical Association}, vol.~103, no.~482, pp.~681--686, 2008.

\bibitem{casella2010penalized}
G.~Casella, M.~Ghosh, J.~Gill, and M.~Kyung, ``Penalized regression, standard
  errors, and bayesian lassos,'' {\em Bayesian Analysis}, vol.~06, no.~2,
  pp.~369--411, 2010.

\bibitem{zou2006adaptive}
H.~Zou, ``The adaptive lasso and its oracle properties,'' {\em Journal of the
  American Statistical Association}, vol.~101, no.~476, pp.~1418--1429, 2006.

\bibitem{karlis2002type}
D.~Karlis, ``An em type algorithm for maximum likelihood estimation of the
  normal--inverse gaussian distribution,'' {\em Statistics \& probability
  letters}, vol.~57, no.~1, pp.~43--52, 2002.

\bibitem{tibshirani1996regression}
R.~Tibshirani, ``Regression shrinkage and selection via the lasso,'' {\em
  Journal of the Royal Statistical Society: Series B (Methodological)},
  vol.~58, no.~1, pp.~267--288, 1996.

\bibitem{wu2009variable}
Y.~Wu and Y.~Liu, ``Variable selection in quantile regression,'' {\em
  Statistica Sinica}, vol.~19, no.~2, pp.~801--817, 2009.

\bibitem{li20081}
Y.~Li and J.~Zhu, ``L 1-norm quantile regression,'' {\em Journal of
  Computational and Graphical Statistics}, vol.~17, no.~1, pp.~163--185, 2008.

\bibitem{wang2009quantile}
H.~J. Wang, Z.~Zhu, and J.~Zhou, ``Quantile regression in partially linear
  varying coefficient models,'' {\em The Annals of Statistics}, pp.~3841--3866,
  2009.

\bibitem{tang2013variable}
Y.~Tang, H.~J. Wang, and Z.~Zhu, ``Variable selection in quantile varying
  coefficient models with longitudinal data,'' {\em Computational Statistics \&
  Data Analysis}, vol.~57, no.~1, pp.~435--449, 2013.

\bibitem{sherwood2023package}
B.~Sherwood, A.~Maidman, M.~B. Sherwood, and T.~ByteCompile, ``Package
  ‘rqpen’,'' {\em Penalized Quantile Regression. In}, vol.~1, 2023.

\bibitem{li2015bayesian}
J.~Li, Z.~Wang, R.~Li, and R.~Wu, ``Bayesian group lasso for nonparametric
  varying-coefficient models with application to functional genome-wide
  association studies,'' {\em The Annals of Applied Statistics}, vol.~9, no.~2,
  p.~640, 2015.

\bibitem{wu2014integrative}
C.~Wu, Y.~Cui, and S.~Ma, ``Integrative analysis of gene--environment
  interactions under a multi-response partially linear varying coefficient
  model,'' {\em Statistics in Medicine}, vol.~33, no.~28, pp.~4988--4998, 2014.

\bibitem{wang2018hydrogen}
D.~Wang, L.~Wang, Y.~Zhang, Y.~Zhao, and G.~Chen, ``Hydrogen gas inhibits lung
  cancer progression through targeting smc3,'' {\em Biomedicine \&
  Pharmacotherapy}, vol.~104, pp.~788--797, 2018.

\bibitem{malik2019micrornas}
S.~Malik, R.~Zafar~Paracha, M.~Khalid, M.~Nisar, A.~Siddiqa, Z.~Hussain,
  R.~Nawaz, A.~Ali, and J.~Ahmad, ``Micrornas and their target mrnas as
  potential biomarkers among smokers and non-smokers with lung
  adenocarcinoma,'' {\em IET Systems Biology}, vol.~13, no.~2, pp.~69--76,
  2019.

\bibitem{mecoli2021cancer}
C.~A. Mecoli, B.~L. Adler, Q.~Yang, L.~K. Hummers, A.~Rosen, L.~Casciola-Rosen,
  and A.~A. Shah, ``Cancer in systemic sclerosis: analysis of antibodies
  against components of the th/to complex,'' {\em Arthritis \& Rheumatology},
  vol.~73, no.~2, pp.~315--323, 2021.

\bibitem{steen2007changes}
V.~D. Steen and T.~A. Medsger, ``Changes in causes of death in systemic
  sclerosis, 1972--2002,'' {\em Annals of the Rheumatic Diseases}, vol.~66,
  no.~7, pp.~940--944, 2007.

\bibitem{naik2021cutaneous}
P.~P. Naik, ``Cutaneous malignant melanoma: A review of early diagnosis and
  management,'' {\em World journal of oncology}, vol.~12, no.~1, p.~7, 2021.

\bibitem{surman2023similarities}
M.~Surman, U.~Jankowska, M.~Wilczak, and M.~Przyby{\l}o, ``Similarities and
  differences in the protein composition of cutaneous melanoma cells and their
  exosomes identified by mass spectrometry,'' {\em Cancers}, vol.~15, no.~4,
  p.~1097, 2023.

\bibitem{orellana2021connexins}
V.~P. Orellana, A.~Tittarelli, and M.~A. Retamal, ``Connexins in melanoma:
  Potential role of cx46 in its aggressiveness,'' {\em Pigment Cell \& Melanoma
  Research}, vol.~34, no.~5, pp.~853--868, 2021.

\bibitem{benjamin2014inositol}
D.~I. Benjamin, S.~M. Louie, M.~M. Mulvihill, R.~A. Kohnz, D.~S. Li, L.~G.
  Chan, A.~Sorrentino, S.~Bandyopadhyay, A.~Cozzo, A.~Ohiri, {\em et~al.},
  ``Inositol phosphate recycling regulates glycolytic and lipid metabolism that
  drives cancer aggressiveness,'' {\em ACS chemical biology}, vol.~9, no.~6,
  pp.~1340--1350, 2014.

\bibitem{angrisani2021emerging}
A.~Angrisani, A.~Di~Fiore, E.~De~Smaele, and M.~Moretti, ``The emerging role of
  the kctd proteins in cancer,'' {\em Cell Communication and Signaling},
  vol.~19, no.~1, p.~56, 2021.

\bibitem{carbonetto2012scalable}
P.~Carbonetto and M.~Stephens, ``Scalable variational inference for bayesian
  variable selection in regression, and its accuracy in genetic association
  studies,'' 2012.

\bibitem{komodromos2022variational}
M.~Komodromos, E.~O. Aboagye, M.~Evangelou, S.~Filippi, and K.~Ray,
  ``Variational bayes for high-dimensional proportional hazards models with
  applications within gene expression,'' {\em Bioinformatics}, vol.~38, no.~16,
  pp.~3918--3926, 2022.

\bibitem{wang2020simple}
G.~Wang, A.~Sarkar, P.~Carbonetto, and M.~Stephens, ``A simple new approach to
  variable selection in regression, with application to genetic fine mapping,''
  {\em Journal of the Royal Statistical Society Series B: Statistical
  Methodology}, vol.~82, no.~5, pp.~1273--1300, 2020.

\bibitem{nie2023bayesian}
L.~Nie and V.~Ro{\v{c}}kov{\'a}, ``Bayesian bootstrap spike-and-slab lasso,''
  {\em Journal of the American Statistical Association}, vol.~118, no.~543,
  pp.~2013--2028, 2023.

\bibitem{fan2008sure}
J.~Fan and J.~Lv, ``Sure independence screening for ultrahigh dimensional
  feature space,'' {\em Journal of the Royal Statistical Society Series B:
  Statistical Methodology}, vol.~70, no.~5, pp.~849--911, 2008.

\bibitem{wang2012quantile}
L.~Wang, Y.~Wu, and R.~Li, ``Quantile regression for analyzing heterogeneity in
  ultra-high dimension,'' {\em Journal of the American Statistical
  Association}, vol.~107, no.~497, pp.~214--222, 2012.

\bibitem{huang2012selective}
J.~Huang, P.~Breheny, and S.~Ma, ``A selective review of group selection in
  high-dimensional models,'' {\em Statistical science: a review journal of the
  Institute of Mathematical Statistics}, vol.~27, no.~4, 2012.

\bibitem{zhou2021gene}
F.~Zhou, J.~Ren, X.~Lu, S.~Ma, and C.~Wu, ``Gene--environment interaction: A
  variable selection perspective,'' {\em Epistasis: Methods and Protocols},
  pp.~191--223, 2021.

\bibitem{koenker1978regression}
R.~Koenker and G.~Bassett~Jr, ``Regression quantiles,'' {\em Econometrica:
  journal of the Econometric Society}, pp.~33--50, 1978.

\bibitem{zou2008composite}
H.~Zou and M.~Yuan, ``Composite quantile regression and the oracle model
  selection theory,'' {\em The Annals of Statistics}, vol.~36, no.~3,
  pp.~1108--1126, 2008.

\bibitem{huang2015bayesian}
H.~Huang and Z.~Chen, ``Bayesian composite quantile regression,'' {\em Journal
  of Statistical Computation and Simulation}, vol.~85, no.~18, pp.~3744--3754,
  2015.

\bibitem{yang2016posterior}
Y.~Yang, H.~J. Wang, and X.~He, ``Posterior inference in bayesian quantile
  regression with asymmetric laplace likelihood,'' {\em International
  Statistical Review}, vol.~84, no.~3, pp.~327--344, 2016.

\bibitem{bodmer2002disruption}
D.~Bodmer, M.~Eleveld, E.~Kater-Baats, I.~Janssen, B.~Janssen, M.~Weterman,
  E.~Schoenmakers, M.~Nickerson, M.~Linehan, B.~Zbar, {\em et~al.},
  ``Disruption of a novel mfs transporter gene, dirc2, by a familial renal cell
  carcinoma-associated t (2; 3)(q35; q21),'' {\em Human molecular genetics},
  vol.~11, no.~6, pp.~641--649, 2002.

\bibitem{savalas2011disrupted}
L.~R.~T. Savalas, B.~Gasnier, M.~Damme, T.~L{\"u}bke, C.~Wrocklage,
  C.~Debacker, A.~J{\'e}z{\'e}gou, T.~Reinheckel, A.~Hasilik, P.~Saftig, {\em
  et~al.}, ``Disrupted in renal carcinoma 2 (dirc2), a novel transporter of the
  lysosomal membrane, is proteolytically processed by cathepsin l,'' {\em
  Biochemical Journal}, vol.~439, no.~1, pp.~113--128, 2011.

\bibitem{deng2019toe1}
T.~Deng, Y.~Huang, K.~Weng, S.~Lin, Y.~Li, G.~Shi, Y.~Chen, J.~Huang, D.~Liu,
  W.~Ma, {\em et~al.}, ``Toe1 acts as a 3' exonuclease for telomerase rna and
  regulates telomere maintenance,'' {\em Nucleic acids research}, vol.~47,
  no.~1, pp.~391--405, 2019.

\bibitem{lardelli2017biallelic}
R.~M. Lardelli, A.~E. Schaffer, V.~R. Eggens, M.~S. Zaki, S.~Grainger,
  S.~Sathe, E.~L. Van~Nostrand, Z.~Schlachetzki, B.~Rosti, N.~Akizu, {\em
  et~al.}, ``Biallelic mutations in the 3' exonuclease toe1 cause
  pontocerebellar hypoplasia and uncover a role in snrna processing,'' {\em
  Nature genetics}, vol.~49, no.~3, pp.~457--464, 2017.

\bibitem{wang2020armcx}
T.~Wang, H.~Zhong, Y.~Qin, W.~Wei, Z.~Li, M.~Huang, X.~Luo, {\em et~al.},
  ``Armcx family gene expression analysis and potential prognostic biomarkers
  for prediction of clinical outcome in patients with gastric carcinoma,'' {\em
  BioMed Research International}, vol.~2020, 2020.

\bibitem{zeller2012candidate}
C.~Zeller, W.~Dai, N.~L. Steele, A.~Siddiq, A.~J. Walley, C.~Wilhelm-Benartzi,
  S.~Rizzo, A.~Van Der~Zee, J.~Plumb, and R.~Brown, ``Candidate dna methylation
  drivers of acquired cisplatin resistance in ovarian cancer identified by
  methylome and expression profiling,'' {\em Oncogene}, vol.~31, no.~42,
  pp.~4567--4576, 2012.

\end{thebibliography}

\clearpage
\section*{Appendix}
\appendix
\section{Details on Alternative Methods}
Alternative methods under comparison are LASSO, quantile LASSO and the spike-and-slab LASSO (ssLASSO). The first two methods have been implemented in R packages \emph{glmnet} \cite{friedman2010regularization} and \emph{rqPen} \cite{sherwood2023package}, respectively. Below we provide the details of ssLASSO. Note that the derivations for the original version of Spike-and-Slab LASSO are avaiable from published studies \cite{rovckova2018spike,tang2017spike,tang2017spikeGLM}. The ssLASSO adopted in this studies slightly differs in including an term corresponding to low-dimensional clinical and environmental variables that are not subject to selection.

\section{The Spike-and-Slab LASSO}
Denote $y_{i}$ as the response of the $i$-th subject $(1\leqslant i\leqslant n)$. Let $\boldsymbol z_{i}=(1,z_{i1},\dots,z_{iq})^{\top}$ be a a $(q+1)$-dimensional vector for the intercept and $q$ clinical factors. In addition, we denote $\boldsymbol x_{i}=(x_{i1},\dots,x_{ip})^{\top}$ as a $p$-dimensional vector for high-dimensional genetic factors. Consider the following linear regression model:
\begin{equation*}
y_i=\boldsymbol z_i^\top\bm{\alpha}+\boldsymbol x_i^\top\bm{\beta}+\epsilon_i,
\end{equation*}
where  coefficient vector $\bm{\alpha}=(\alpha_0,\cdots,\alpha_q)^\top$ represents regression coefficients for the intercept and clinical covariates, respectively. The coefficient vector  $\bm{\beta}=(\beta_1,\cdots,\beta_p)^\top$ is corresponding to the $p$ dimensional omics features. We assume that the random error $\epsilon_i$ follows a normal distribution with standard deviation $\tilde{\sigma}$.

The spike-and-slab mixture double-exponential prior is  assigned on each component of $\bm{\beta}$ as:
\begin{equation*}
\beta_j|\gamma_j,s_0,s_1\overset{ind}{\sim}(1-\gamma_j)\pi(\beta_j|s_0)+\gamma_j\pi(\beta_j|s_1),
\end{equation*}
with $\gamma_j=0$ or $1$ indicating the corresponding coefficient $\beta_j$ being $0$ or not. With $s_1>s_0>0$, the scale parameter $s_0$ controls the spike part which models small and negligible coefficients that are  close to $0$,  and $s_1$ controls the diffuse slab distribution which models important effects corresponding to regression coefficients with large magnitude. The above prior can also be viewed as $\beta_j|S_j\overset{ind}{\sim} \frac{1}{2S_j}\text{exp}\left( -\frac{|\beta_j|}{S_j}\right)$, where  $S_j=(1-\gamma_j)s_0+\gamma_js_1$. Regular double exponential prior is a special case when $s_0=s_1$ or $\gamma_j$ is set to be $0$ or $1$.

The indicator variable $\gamma_j$ is assumed to independently follow a Bernoulli distribution:
\begin{equation*}
\gamma_j|\theta\overset{ind}{\sim}\text{Ber}(\theta)=\theta^{\gamma_j}(1-\theta)^{1-\gamma_j},
\end{equation*}
with the probability parameter $\theta$ being assigned a uniform prior $\theta\sim U(0,1)$. It can be viewed as probability $p(\gamma_j=1|\theta)$, a global shrinkage parameter indicating the prior probability of nonzero components in $\bm{\beta}$. Estimating different $S_j$s will lead to different level of regularization for each coefficient.

The intercept and each clinical factor $\bm{\alpha}_k$ are assumed to have independent normal priors with mean 0 and variance $V_k$:
\begin{equation*}
\alpha_k\overset{ind}{\sim}\text{N}(0,V_k),
\end{equation*}
where $V_k$ is set to $10^3$ for computational convenience.
For variance $\tilde{\sigma}^2$, we adopt the limiting improper prior:
\begin{equation*}
\pi(\tilde{\sigma}^2) =\tilde{\sigma}^{-2}. 
\end{equation*}
Recall that $S_j=(1-\gamma_j)s_0+\gamma_j s_1$, and denote $\boldsymbol \gamma = (\gamma_1,...,\gamma_p)^\top$. After omitting the irrelevant terms, the log joint posterior density function is given by
\begin{equation}\label{logBLSS}
\begin{split}
Q&=\text{log}p(\bm{\alpha},\bm{\beta},\tilde{\sigma}^2,\boldsymbol \gamma,\theta|\boldsymbol Y)\\
&=\text{log}p(\boldsymbol Y|\bm{\alpha},\bm{\beta},\tilde{\sigma}^2)+\sum_{k=0}^q\text{log}p(\alpha_k)+\sum_{j=1}^{p}\text{log} p(\beta_j|S_j)+\text{log}p(\tilde{\sigma}^2)+\sum_{j=1}^{p}\text{log} p(\gamma_j|\theta)+\text{log} p(\theta)\\
&\propto -\frac{n}{2}\text{log}(\tilde{\sigma}^2)-\frac{1}{2\tilde{\sigma}^2}\sum_{i=1}^n(y_i-\boldsymbol z_i^\top\bm{\alpha}-\boldsymbol x_i^\top\bm{\beta})^2-\sum_{k=0}^q\frac{\alpha_k^2}{2V_k}-\sum_{j=1}^{p}\frac{1}{S_j}|\beta_j |-\text{log}(\tilde{\sigma}^2)\\
&\ \ \ +\sum_{j=1}^{p}\left[\gamma_j \text{log}\theta+(1-\gamma_j)\text{log}(1-\theta)\right].
\end{split}
\end{equation}
At E-step, we calculate the conditional expectations of $\gamma_j$ at $d$-th iteration:
\begin{equation}\label{gammaj2}
\begin{split}
\eta_j^{(d)}&=p(\gamma_j=1|\beta_j^{(d)},\theta^{(d)},\boldsymbol Y)\\
&=\frac{p(\beta_j^{(d)}|\gamma_j=1,s_1)p(\gamma_j=1|\theta^{(d)})}{p(\beta_j^{(d)}|\gamma_j=0,s_0)p(\gamma_j=0|\theta^{(d)})+p(\beta_j^{(d)}|\gamma_j=1,s_1)p(\gamma_j=1|\theta^{(d)})}\\
&=\frac{\pi(\beta_j^{(d)}|s_1)\theta^{(d)}}{\pi(\beta_j^{(d)}|s_0)(1-\theta^{(d)})+\pi(\beta_j^{(d)}|s_1)\theta^{(d)}}.
\end{split}
\end{equation}
Therefore, the conditional expectation of $(S_j^{-1})^{(d)}$ can be computed as: 
\begin{equation}\label{Sj2}
(\tilde{S}_j^{-1})^{(d)}=E((S_j^{-1})^{(d)}|\beta_j^{(d)})=E\left(\frac{1}{(1-\gamma_j)s_0+\gamma_j s_1}\middle|\beta_j^{(d)}\right)=\frac{1-\eta_j^{(d)}}{s_0}+\frac{\eta_j^{(d)}}{s_1}.
\end{equation}
At M-step, we update $\bm{\alpha}^{(d+1)}$, $\bm{\beta}^{(d+1)}$, $(\tilde{\sigma}^2)^{(d+1)}$ and $\theta^{(d+1)}$ by maximizing the log joint posterior density at the $d$-th iteration, or $Q(\bm{\phi}|\bm{\phi}^{(d)})$ where $\bm{\phi}=(\bm{\alpha},\bm{\beta},\tilde{\sigma}^2,\boldsymbol \gamma,\theta)$. In addition, we update the coefficient vector $\bm{\alpha}$ and $\bm{\beta}$ component-wisely by following the coordinate descent strategy \cite{friedman2010regularization}. By solving $\frac{\partial Q(\bm{\phi}|\bm{\phi}^{(d)})}{\partial \alpha_l}=0$, $\frac{\partial Q(\bm{\phi}|\bm{\phi}^{(d)})}{\partial \beta_m}=0$, $\frac{\partial Q(\bm{\phi}|\bm{\phi}^{(d)})}{\partial \tilde{\sigma}^2}=0$ and $\frac{\partial Q(\bm{\phi}|\bm{\phi}^{(d)})}{\partial \theta}=0$, , where $l=1,...,q$ and $m=1,...,p$, we obtain the following equations:

\begin{equation}\label{cd2}
\begin{split}
\theta^{(d+1)}&=\frac{1}{p}\sum_{j=1}^{p}\eta_j^{(d)},\\
(\tilde{\sigma}^2)^{(d+1)}&=\frac{\sum_{i=1}^n(y_i-\boldsymbol z_i^\top\bm{\alpha}^{(d)}-\boldsymbol x_i^\top\bm{\beta}^{(d)})^2}{n+2},\\
\alpha_l^{(d+1)}&=\frac{\sum_{i=1}^{n}\left( y_i-\boldsymbol x_i^\top\bm{\beta}^{(d)}-\sum_{k=0}^{l-1}z_{ik}\alpha_k^{(d+1)}-\sum_{k=l+1}^{q}z_{ik}\alpha_k^{(d)}\right)z_{il}}{\frac{2(\tilde{\sigma}^2)^{(d+1)}}{V_l}+\sum_{i=1}^{n}z_{il}^2},\\
\beta_m^{(d+1)}&=\frac{\text{sgn}(T_m^{(d+1)})\left(\left| T_m^{(d+1)}\right|-(\tilde{S}_m^{-1})^{(d)}\right)_+}{\sum_{i=1}^nx_{im}^2/(\tilde{\sigma}^2)^{(d+1)}},
\end{split}
\end{equation}
where
\begin{equation*}
T_m^{(d+1)}=\frac{1}{(\tilde{\sigma}^2)^{(d+1)}}\sum_{i=1}^n\left(y_i-\sum_{j=1}^{m-1}x_{ij}\beta_j^{(d+1)}-\sum_{j=m+1}^{p}x_{ij}\beta_j^{(d)}-\boldsymbol z_i^\top\bm{\alpha}^{(d+1)} \right)x_{im}.
\end{equation*}
We do not perform variable selection on the clinical factors. Thus, shrinkage is only imposed on $\beta$. A $k$-fold cross-validation is used to choose the optimal tuning parameters $s_0$ and $s_1$. For fixed tuning parameters, the model fitting algorithm is provided in the table below:






\begin{table} [hbt!]
	\def\arraystretch{1.3}
	\begin{center}
		\caption{EM algorithm for the spike-and-slab LASSO.}\label{alg1}
		\centering
		\fontsize{10}{10}\selectfont{
\begin{tabularx}{\textwidth}{ m{0.9\textwidth} }
				\hline
				\vspace{0.05cm}
				\textbf{Algorithm} \\[0.1ex]
				\hline
				\vspace{0.1cm}
				\hspace{1em} Initialize iteration $d = 0$, coefficients $\bm{\alpha}^{(0)}$, $\bm{\beta}^{(0)}$, $\theta^{(0)}$ and $(\tilde{\sigma}^2)^{(0)}$;\\
				\hspace{1em} \textbf{Repeat} \\
				\hspace{3em} \textbf{E-step:}\\
				\hspace{3em} Compute the conditional expectation of $\gamma_j^{(d)}$ and $(\tilde{S}_j^{-1})^{(d)}$ via (\ref{gammaj2}) and (\ref{Sj2});\\
				\hspace{3em} \textbf{M-step:}\\
				\hspace{3em} Update $\theta^{(d+1)}$, $(\tilde{\sigma}^2)^{(d+1)}$, $\alpha_l^{(d+1)}$ and $\beta_m^{(d+1)}$ through (\ref{cd2}); \\
				\hspace{3em} Calculate the updated log joint posterior density and check if $|Q^{(d+1)}-Q^{(d)}|<\delta$;\\
				\hspace{3em} $d\leftarrow d +1$; \\
				\hspace{1em} \textbf{until} convergence ($|Q^{(d+1)}-Q^{(d)}|<\delta$ for a small value $\delta>0$).\\
				\hline
		\end{tabularx} }
	\end{center}
	\centering
\end{table}

\section{Additional Simulations}

All the tables listed in this Section have been referred in the Section of Simulation.

\begin{table} [H]
	\def\arraystretch{1.5}
	\captionsetup{font=scriptsize}
\begin{center}
    \caption{Evaluation of homogeneous model under AR-1 correlation, $(n,p)$=(400,800) with 100 replicates.}\label{sim1}
    \centering
		\fontsize{7}{7}\selectfont{
\begin{tabular}{ l l l l l l l l }
 \hline
 $n=400$&$p=800$&& \multicolumn{1}{c}{TP}& \multicolumn{1}{c}{FP} & \multicolumn{1}{c}{F1} & \multicolumn{1}{c}{MCC} & \multicolumn{1}{c}{Estimation}\\
 \hline

$\tau=0.3$& Error1 & ssQLASSO & 15.00(0.00) & 1.53(2.03) & 0.95(0.06) & 0.96(0.05) & 0.81(0.18) \\ 
   &  & QLASSO & 15.00(0.00) & 19.47(4.44) & 0.61(0.06) & 0.66(0.05) & 3.11(0.28) \\ 
   &  & ssLASSO & 15.00(0.00) & 0.23(0.63) & 0.99(0.02) & 0.99(0.02) & 0.63(0.17) \\ 
   &  & LASSO & 15.00(0.00) & 58.40(24.56) & 0.36(0.09) & 0.45(0.07) & 2.97(0.72) \\
\cline{2-8}

& Error2 & ssQLASSO & 14.90(0.40) & 0.87(1.11) & 0.97(0.04) & 0.97(0.03) & 2.53(1.30) \\ 
   &  & QLASSO & 15.00(0.00) & 20.03(4.55) & 0.60(0.05) & 0.65(0.04) & 4.93(0.85) \\ 
   &  & ssLASSO & 10.03(3.27) & 0.23(0.50) & 0.77(0.17) & 0.79(0.14) & 4.60(1.96) \\
\hline
\end{tabular}
}
\end{center}
\centering
\end{table}

\begin{table} [H]
	\def\arraystretch{1.5}
	\captionsetup{font=scriptsize}
\begin{center}
    \caption{Evaluation of homogeneous model under banded correlation, $(n,p)$=(400,800) with 100 replicates.}\label{sim3}
    \centering
		\fontsize{7}{7}\selectfont{
\begin{tabular}{ l l l l l l l l }
 \hline
 $n=400$&$p=800$&& \multicolumn{1}{c}{TP}& \multicolumn{1}{c}{FP} & \multicolumn{1}{c}{F1} & \multicolumn{1}{c}{MCC} & \multicolumn{1}{c}{Estimation}\\
 \hline

$\tau=0.3$& Error1 & ssQLASSO & 15.00(0.00) & 0.80(1.45) & 0.98(0.04) & 0.98(0.04) & 0.89(0.21) \\ 
   &  & QLASSO & 15.00(0.00) & 18.47(5.62) & 0.63(0.07) & 0.67(0.06) & 2.95(0.37) \\ 
   &  & ssLASSO & 15.00(0.00) & 0.47(0.78) & 0.99(0.02) & 0.99(0.02) & 0.71(0.20) \\ 
   &  & LASSO & 15.00(0.00) & 59.03(23.25) & 0.36(0.09) & 0.45(0.08) & 3.00(0.67) \\
\cline{2-8}
& Error2 & ssQLASSO & 14.83(0.38) & 0.50(0.78) & 0.98(0.02) & 0.98(0.02) & 2.43(0.99) \\ 
   &  & QLASSO & 14.97(0.18) & 19.77(4.12) & 0.61(0.05) & 0.65(0.04) & 4.69(0.80) \\ 
   &  & ssLASSO & 10.63(2.66) & 0.10(0.31) & 0.81(0.13) & 0.83(0.11) & 4.24(1.64) \\
\hline
\end{tabular}
}
\end{center}
\centering
\end{table}

\begin{table} [H]
	\def\arraystretch{1.5}
	\captionsetup{font=scriptsize}
\begin{center}
    \caption{Evaluation of homogeneous model under banded correlation, $(n,p)$=(400,1600) with 100 replicates.}\label{sim4}
    \centering
		\fontsize{7}{7}\selectfont{
\begin{tabular}{ l l l l l l l l }
 \hline
 $n=400$&$p=1600$&& \multicolumn{1}{c}{TP}& \multicolumn{1}{c}{FP} & \multicolumn{1}{c}{F1} & \multicolumn{1}{c}{MCC} & \multicolumn{1}{c}{Estimation}\\
 \hline

$\tau=0.3$& Error1 & ssQLASSO & 15.00(0.00) & 1.77(2.92) & 0.95(0.08) & 0.95(0.07) & 0.86(0.18) \\ 
   &  & QLASSO & 15.00(0.00) & 37.20(7.61) & 0.45(0.05) & 0.53(0.04) & 3.31(0.49) \\ 
   &  & ssLASSO & 15.00(0.00) & 0.87(1.04) & 0.97(0.03) & 0.97(0.03) & 0.69(0.18) \\ 
   &  & LASSO & 15.00(0.00) & 69.07(22.24) & 0.32(0.07) & 0.42(0.06) & 3.38(0.69) \\
\cline{2-8}
& Error2 & ssQLASSO & 14.60(0.89) & 1.23(2.25) & 0.95(0.06) & 0.95(0.06) & 3.39(1.41) \\ 
   &  & QLASSO & 15.00(0.00) & 32.33(5.77) & 0.49(0.05) & 0.56(0.04) & 5.60(1.09) \\ 
   &  & ssLASSO & 8.57(3.65) & 0.27(0.58) & 0.69(0.21) & 0.72(0.18) & 5.68(2.35) \\
\hline
\end{tabular}
}
\end{center}
\centering
\end{table}

\begin{table} [H]
	\def\arraystretch{1.5}
	\captionsetup{font=scriptsize}
\begin{center}
    \caption{Evaluation of heterogeneous model under AR-1 correlation, $(n,p)$=(400,800) with 100 replicates.}\label{sim5}
    \centering
		\fontsize{7}{7}\selectfont{
\begin{tabular}{ l l l l l l l l }
 \hline
 $n=400$&$p=800$&& \multicolumn{1}{c}{TP}& \multicolumn{1}{c}{FP} & \multicolumn{1}{c}{F1} & \multicolumn{1}{c}{MCC} & \multicolumn{1}{c}{Estimation}\\
 \hline

$\tau=0.3$& Error1 & ssQLASSO & 15.00(0.00) & 0.63(1.16) & 0.98(0.03) & 0.98(0.03) & 1.12(0.12) \\ 
   &  & QLASSO & 15.00(0.00) & 18.70(5.13) & 0.62(0.06) & 0.66(0.05) & 3.22(0.47) \\ 
   &  & ssLASSO & 15.00(0.00) & 0.07(0.25) & 1.00(0.01) & 1.00(0.01) & 1.39(0.22) \\ 
   &  & LASSO & 15.00(0.00) & 58.77(21.41) & 0.35(0.08) & 0.45(0.07) & 4.46(1.07) \\
\cline{2-8}
& Error2 & ssQLASSO & 14.97(0.18) & 0.30(0.99) & 0.99(0.03) & 0.99(0.03) & 1.27(0.59) \\ 
   &  & QLASSO & 14.97(0.18) & 19.27(3.77) & 0.61(0.05) & 0.66(0.04) & 4.35(1.21) \\ 
   &  & ssLASSO & 12.07(2.84) & 0.20(0.76) & 0.87(0.14) & 0.88(0.12) & 4.39(2.29) \\
\hline
\end{tabular}
}
\end{center}
\centering
\end{table}

\begin{table} [H]
	\def\arraystretch{1.5}
	\captionsetup{font=scriptsize}
\begin{center}
    \caption{Evaluation of heterogeneous model under AR-1 correlation, $(n,p)$=(400,1600) with 100 replicates.}\label{sim6}
    \centering
		\fontsize{7}{7}\selectfont{
\begin{tabular}{ l l l l l l l l }
 \hline
 $n=400$&$p=1600$&& \multicolumn{1}{c}{TP}& \multicolumn{1}{c}{FP} & \multicolumn{1}{c}{F1} & \multicolumn{1}{c}{MCC} & \multicolumn{1}{c}{Estimation}\\
 \hline

$\tau=0.3$& Error1 & ssQLASSO & 15.00(0.00) & 0.87(1.50) & 0.97(0.04) & 0.97(0.04) & 1.07(0.18) \\ 
   &  & QLASSO & 15.00(0.00) & 34.53(7.38) & 0.47(0.06) & 0.55(0.05) & 3.39(0.39) \\ 
   &  & ssLASSO & 15.00(0.00) & 0.07(0.37) & 1.00(0.01) & 1.00(0.01) & 1.26(0.20) \\ 
   &  & LASSO & 15.00(0.00) & 67.17(21.86) & 0.32(0.07) & 0.43(0.06) & 4.61(0.78) \\
\cline{2-8}
& Error2 & ssQLASSO & 14.10(1.75) & 0.23(0.43) & 0.96(0.07) & 0.96(0.07) & 1.89(1.33) \\ 
   &  & QLASSO & 15.00(0.00) & 33.00(4.53) & 0.48(0.03) & 0.55(0.03) & 5.29(1.02) \\ 
   &  & ssLASSO & 6.90(2.81) & 0.07(0.25) & 0.61(0.17) & 0.66(0.14) & 7.10(1.56) \\
\hline
\end{tabular}
}
\end{center}
\centering
\end{table}

\begin{table} [H]
	\def\arraystretch{1.5}
	\captionsetup{font=scriptsize}
\begin{center}
    \caption{Evaluation of heterogeneous model under banded correlation, $(n,p)$=(400,800) with 100 replicates.}\label{sim7}
    \centering
		\fontsize{7}{7}\selectfont{
\begin{tabular}{ l l l l l l l l }
 \hline
 $n=400$&$p=800$&& \multicolumn{1}{c}{TP}& \multicolumn{1}{c}{FP} & \multicolumn{1}{c}{F1} & \multicolumn{1}{c}{MCC} & \multicolumn{1}{c}{Estimation}\\
 \hline

$\tau=0.3$& Error1 & ssQLASSO & 15.00(0.00) & 0.53(0.73) & 0.98(0.02) & 0.98(0.02) & 1.14(0.20) \\ 
   &  & QLASSO & 15.00(0.00) & 17.27(3.90) & 0.64(0.05) & 0.68(0.04) & 3.28(0.48) \\ 
   &  & ssLASSO & 15.00(0.00) & 0.03(0.18) & 1.00(0.01) & 1.00(0.01) & 1.38(0.24) \\ 
   &  & LASSO & 15.00(0.00) & 67.00(17.35) & 0.32(0.06) & 0.42(0.05) & 4.71(0.90) \\
\cline{2-8}
& Error2 & ssQLASSO & 14.97(0.18) & 0.27(0.78) & 0.99(0.02) & 0.99(0.02) & 1.24(0.38) \\ 
   &  & QLASSO & 15.00(0.00) & 19.10(4.48) & 0.62(0.05) & 0.66(0.04) & 4.68(0.83) \\ 
   &  & ssLASSO & 11.50(3.10) & 0.30(0.47) & 0.84(0.15) & 0.85(0.13) & 4.51(1.76) \\
\hline
\end{tabular}
}
\end{center}
\centering
\end{table}

\begin{table} [H]
	\def\arraystretch{1.5}
	\captionsetup{font=scriptsize}
\begin{center}
    \caption{Evaluation of heterogeneous model under banded correlation, $(n,p)$=(400,1600) with 100 replicates.}\label{sim8}
    \centering
		\fontsize{7}{7}\selectfont{
\begin{tabular}{ l l l l l l l l }
 \hline
 $n=400$&$p=1600$&& \multicolumn{1}{c}{TP}& \multicolumn{1}{c}{FP} & \multicolumn{1}{c}{F1} & \multicolumn{1}{c}{MCC} & \multicolumn{1}{c}{Estimation}\\
 \hline

$\tau=0.3$& Error1 & ssQLASSO & 15.00(0.00) & 0.73(1.17) & 0.98(0.04) & 0.98(0.03) & 1.07(0.26) \\ 
   &  & QLASSO & 15.00(0.00) & 33.73(6.06) & 0.47(0.05) & 0.55(0.04) & 3.47(0.48) \\ 
   &  & ssLASSO & 15.00(0.00) & 0.07(0.25) & 1.00(0.01) & 1.00(0.01) & 1.27(0.22) \\ 
   &  & LASSO & 15.00(0.00) & 71.73(21.72) & 0.31(0.07) & 0.42(0.06) & 5.02(0.93) \\
\cline{2-8}
& Error2 & ssQLASSO & 13.80(1.75) & 0.10(0.31) & 0.95(0.07) & 0.95(0.06) & 2.01(1.32) \\ 
   &  & QLASSO & 14.97(0.18) & 32.93(4.49) & 0.48(0.03) & 0.55(0.03) & 5.34(1.01) \\ 
   &  & ssLASSO & 6.60(2.81) & 0.00(0.00) & 0.59(0.18) & 0.65(0.15) & 7.21(1.66) \\
\hline
\end{tabular}
}
\end{center}
\centering
\end{table}

\begin{table} [H]
	\def\arraystretch{1.5}
	\captionsetup{font=scriptsize}
\begin{center}
    \caption{Evaluation of homogeneous model under AR-1 correlation with 8 positive and 7 negative coefficients, $(n,p)$=(400,1600).}\label{sim9}
    \centering
		\fontsize{7}{7}\selectfont{
\begin{tabular}{ l l l l l l l l }
 \hline
 $n=400$&$p=1600$&& \multicolumn{1}{c}{TP}& \multicolumn{1}{c}{FP} & \multicolumn{1}{c}{F1} & \multicolumn{1}{c}{MCC} & \multicolumn{1}{c}{Estimation}\\
 \hline

$\tau=0.3$& Error1 & ssQLASSO & 15.00(0.00) & 1.27(1.91) & 0.96(0.05) & 0.96(0.05) & 0.89(0.22) \\ 
   &  & QLASSO & 15.00(0.00) & 34.20(6.01) & 0.47(0.04) & 0.55(0.03) & 3.27(0.39) \\ 
   &  & ssLASSO & 15.00(0.00) & 0.40(0.86) & 0.99(0.03) & 0.99(0.03) & 0.64(0.12) \\ 
   &  & LASSO & 15.00(0.00) & 71.40(30.91) & 0.32(0.09) & 0.42(0.07) & 3.39(0.84) \\
\cline{2-8}
& Error2 & ssQLASSO & 14.40(1.10) & 0.70(1.12) & 0.96(0.05) & 0.96(0.05) & 3.42(1.36) \\ 
   &  & QLASSO & 15.00(0.00) & 34.13(3.82) & 0.47(0.03) & 0.55(0.02) & 5.82(0.98) \\ 
   &  & ssLASSO & 7.63(3.55) & 0.17(0.38) & 0.64(0.22) & 0.68(0.18) & 6.12(2.20) \\
\hline
\end{tabular}
}
\end{center}
\centering
\end{table}

\begin{table} [H]
	\def\arraystretch{1.5}
	\captionsetup{font=scriptsize}
\begin{center}
    \caption{Evaluation of homogeneous model under real data correlation, $(n,p)$=(400,1600) with 100 replicates.}\label{sim10}
    \centering
		\fontsize{7}{7}\selectfont{
\begin{tabular}{ l l l l l l l l }
 \hline
 $n=400$&$p=1600$&& \multicolumn{1}{c}{TP}& \multicolumn{1}{c}{FP} & \multicolumn{1}{c}{F1} & \multicolumn{1}{c}{MCC} & \multicolumn{1}{c}{Estimation}\\
 \hline

$\tau=0.3$& Error1 & ssQLASSO & 15.00(0.00) & 3.28(1.93) & 0.90(0.05) & 0.91(0.05) & 1.84(0.59) \\ 
   &  & QLASSO & 14.50(0.76) & 25.75(6.94) & 0.53(0.07) & 0.59(0.07) & 6.85(1.25) \\ 
   &  & ssLASSO & 4.67(2.20) & 0.44(0.51) & 0.45(0.16) & 0.52(0.13) & 8.35(1.32) \\ 
   &  & LASSO & 15.00(0.00) & 57.40(9.60) & 0.34(0.04) & 0.44(0.03) & 4.55(0.65) \\ 
\cline{2-8}
& Error2 & ssQLASSO & 12.78(1.80) & 3.28(1.23) & 0.82(0.08) & 0.82(0.08) & 4.35(1.57) \\ 
   &  & QLASSO & 11.90(1.55) & 23.65(4.39) & 0.47(0.06) & 0.50(0.07) & 8.52(1.22) \\ 
   &  & ssLASSO & 4.78(2.69) & 0.94(0.94) & 0.44(0.19) & 0.49(0.16) & 8.58(1.19) \\
\hline
\end{tabular}
}
\end{center}
\centering
\end{table}

\begin{table} [H]
	\def\arraystretch{1.5}
	\captionsetup{font=scriptsize}
\begin{center}
    \caption{Sensitivity evaluation of homogeneous model under AR-1 correlation, $(n,p)$=(400,1600) with 100 replicates.}\label{sim11}
    \centering
		\fontsize{7}{7}\selectfont{
\begin{tabular}{ l l l l l l l l }
 \hline
 $n=400$&$p=1600$&& \multicolumn{1}{c}{TP}& \multicolumn{1}{c}{FP} & \multicolumn{1}{c}{F1} & \multicolumn{1}{c}{MCC} & \multicolumn{1}{c}{Estimation}\\
 \hline

$\tau=0.3$& Error1 & ssQLASSO & 15.00(0.00) & 1.70(2.09) & 0.95(0.06) & 0.95(0.05) & 0.83(0.15) \\ 
$(a,b)=(1,5)$   &  & QLASSO & 15.00(0.00) & 35.33(7.75) & 0.47(0.05) & 0.54(0.04) & 3.27(0.38) \\ 
   &  & ssLASSO & 15.00(0.00) & 0.67(0.99) & 0.98(0.03) & 0.98(0.03) & 0.65(0.20) \\ 
   &  & LASSO & 15.00(0.00) & 75.33(23.92) & 0.30(0.08) & 0.41(0.06) & 3.44(0.69) \\
\cline{2-8}
& Error2 & ssQLASSO & 14.93(0.37) & 0.27(0.78) & 0.99(0.03) & 0.99(0.03) & 1.09(0.36) \\ 
   &  & QLASSO & 15.00(0.00) & 32.03(3.51) & 0.49(0.03) & 0.56(0.02) & 5.38(0.61) \\ 
   &  & ssLASSO & 8.60(3.47) & 0.03(0.18) & 0.70(0.20) & 0.74(0.17) & 5.52(2.20) \\
\hline
\end{tabular}
}
\end{center}
\centering
\end{table}

\begin{table} [H]
	\def\arraystretch{1.5}
	\captionsetup{font=scriptsize}
\begin{center}
    \caption{Sensitivity evaluation of heterogeneous model under AR-1 correlation, $(n,p)$=(400,1600) with 100 replicates.}\label{sim12}
    \centering
		\fontsize{7}{7}\selectfont{
\begin{tabular}{ l l l l l l l l }
 \hline
 $n=400$&$p=1600$&& \multicolumn{1}{c}{TP}& \multicolumn{1}{c}{FP} & \multicolumn{1}{c}{F1} & \multicolumn{1}{c}{MCC} & \multicolumn{1}{c}{Estimation}\\
 \hline

$\tau=0.3$& Error1 & ssQLASSO & 15.00(0.00) & 0.77(0.94) & 0.98(0.03) & 0.98(0.03) & 1.07(0.19) \\ 
$(a,b)=(1,5)$   &  & QLASSO & 15.00(0.00) & 35.53(6.74) & 0.46(0.05) & 0.54(0.04) & 3.70(0.59) \\ 
   &  & ssLASSO & 15.00(0.00) & 0.10(0.40) & 1.00(0.01) & 1.00(0.01) & 1.31(0.20) \\ 
   &  & LASSO & 15.00(0.00) & 76.83(30.14) & 0.30(0.09) & 0.41(0.08) & 5.34(1.25) \\
\cline{2-8}
& Error2 & ssQLASSO & 13.47(2.06) & 0.33(0.66) & 0.93(0.09) & 0.93(0.08) & 2.33(1.51) \\ 
   &  & QLASSO & 14.97(0.18) & 35.43(4.57) & 0.46(0.03) & 0.54(0.03) & 5.52(0.94) \\ 
   &  & ssLASSO & 6.80(2.92) & 0.47(1.66) & 0.59(0.17) & 0.64(0.14) & 7.32(1.69) \\
\hline
\end{tabular}
}
\end{center}
\centering
\end{table}
\FloatBarrier

\newpage
\section{Tuning Selection}

In the numeric study, the tuning parameters of ssQLASSO have been chosen using the Schwarz Information Criterion (SIC). We have also explored the selection of tuning parameter ($s_0$,$s_1$) under cross validation (CV) in terms of the quantile check loss. 


 \begin{figure}[H]
	\begin{center}
		\includegraphics[height=0.3\textheight]{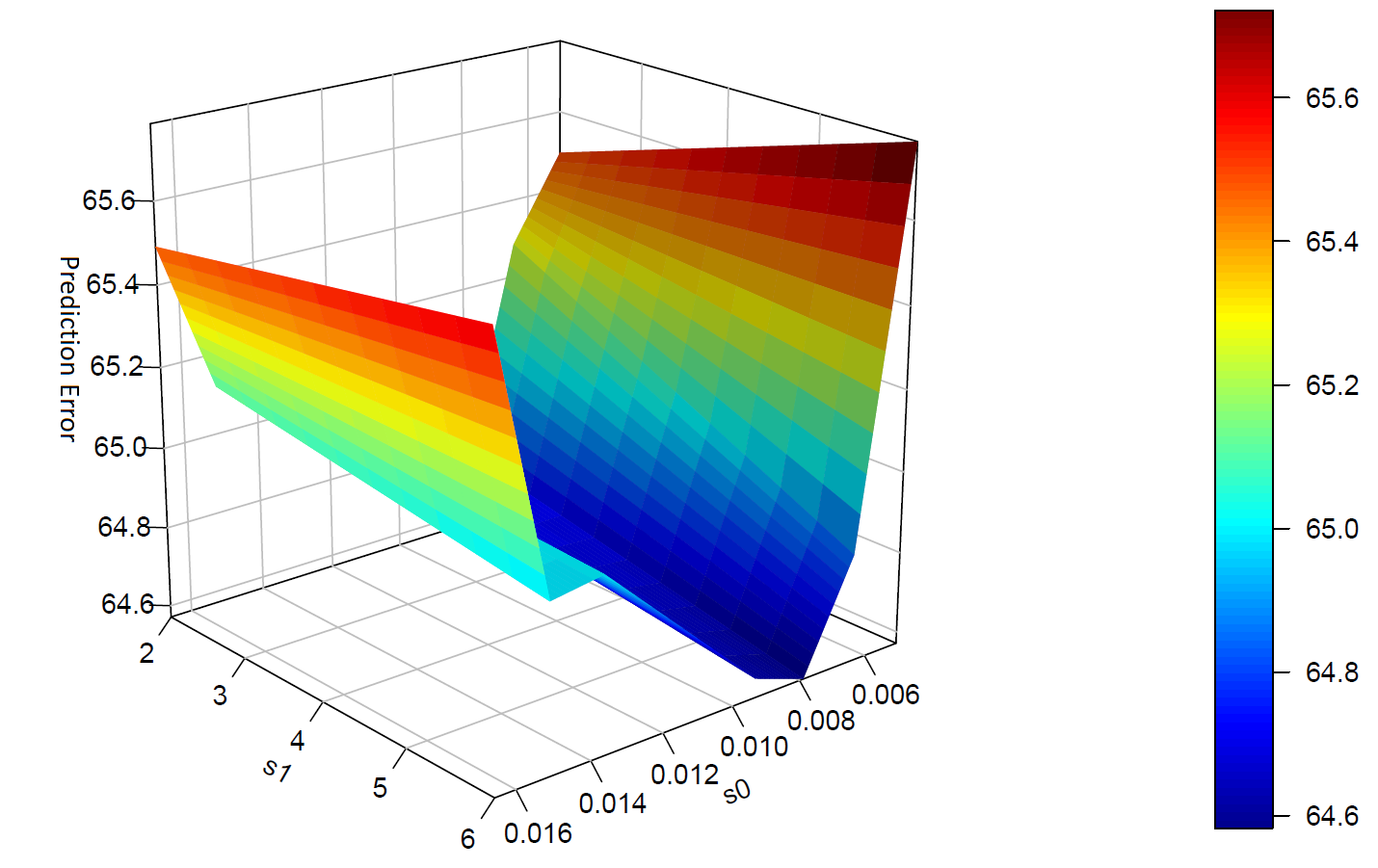}
		\caption{\footnotesize The surface plot of prediction error using ssQLASSO based on quantile check loss under $\tau=70\%$, AR-1 correlation, $(n=400,p=1600)$ and error 3 over 100 replications.}\label{3d}
	\end{center} 
\end{figure}

Figure \ref{3d} shows the 3D surface plot of CV error when ssQLASSO is adopted for model fitting. The pattern is similar to that in Figure \ref{SIC} when best tunings are selected using SIC. Please refer to the subsection of \textbf{Tuning parameter selection} in the Section of Simulation for more details of the interpretation.

\begin{figure}[H]
	\begin{center}
		\includegraphics[height=0.35\textheight]{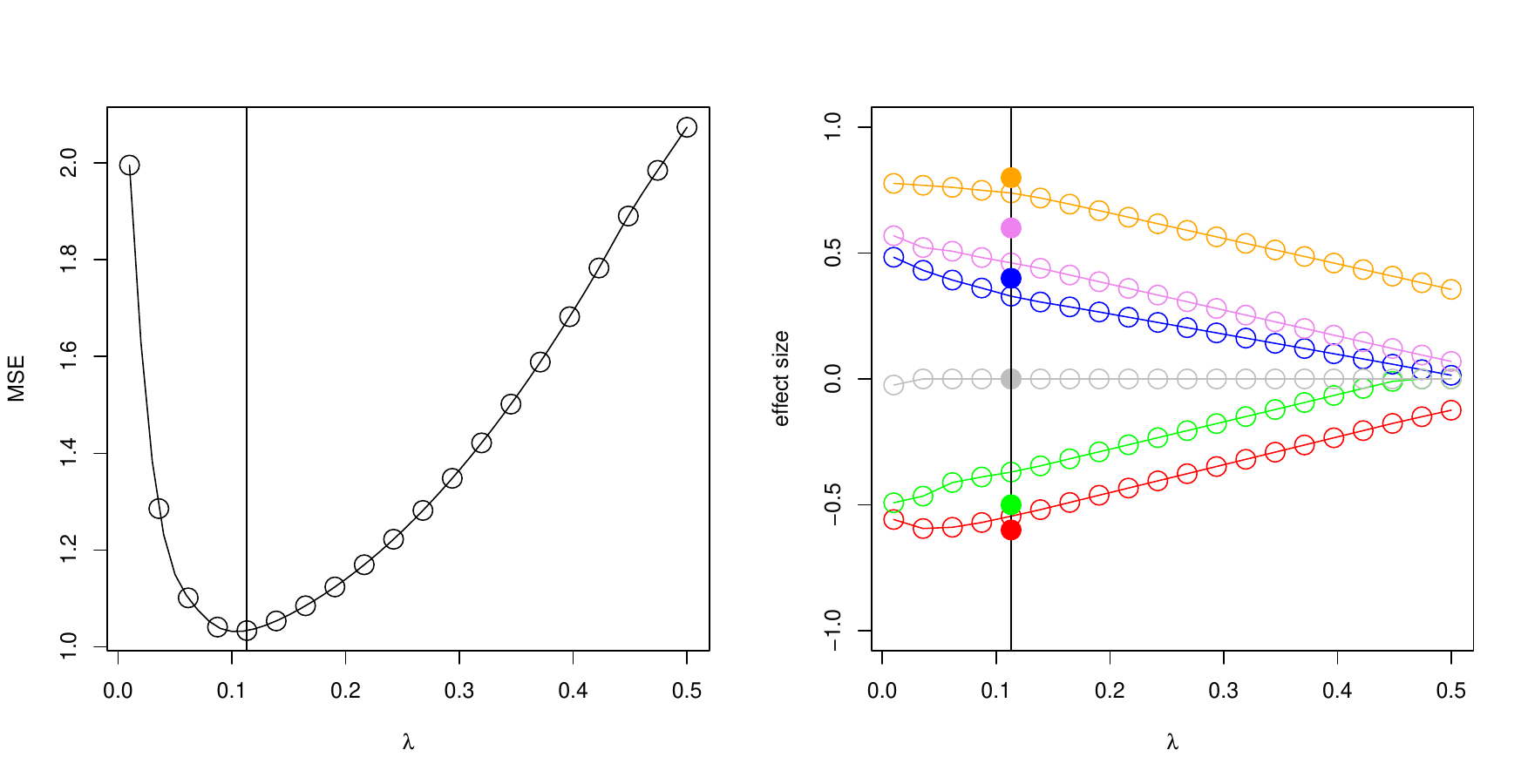}
		\caption{\footnotesize The MSE profile (left) and solution path (right) of LASSO under $\tau=50\%$, AR-1 correlation, $(n=400,p=1600)$ and Error3. The colored circles on the solution path denote the estimated values with respect to the 5 non-zero coefficients represented by filled circles. The vertical lines correspond to the optimal tuning parameter.}\label{LASSOSP}
	\end{center}
\end{figure}
\begin{figure}[H]
	\begin{center}
		\includegraphics[height=0.35\textheight]{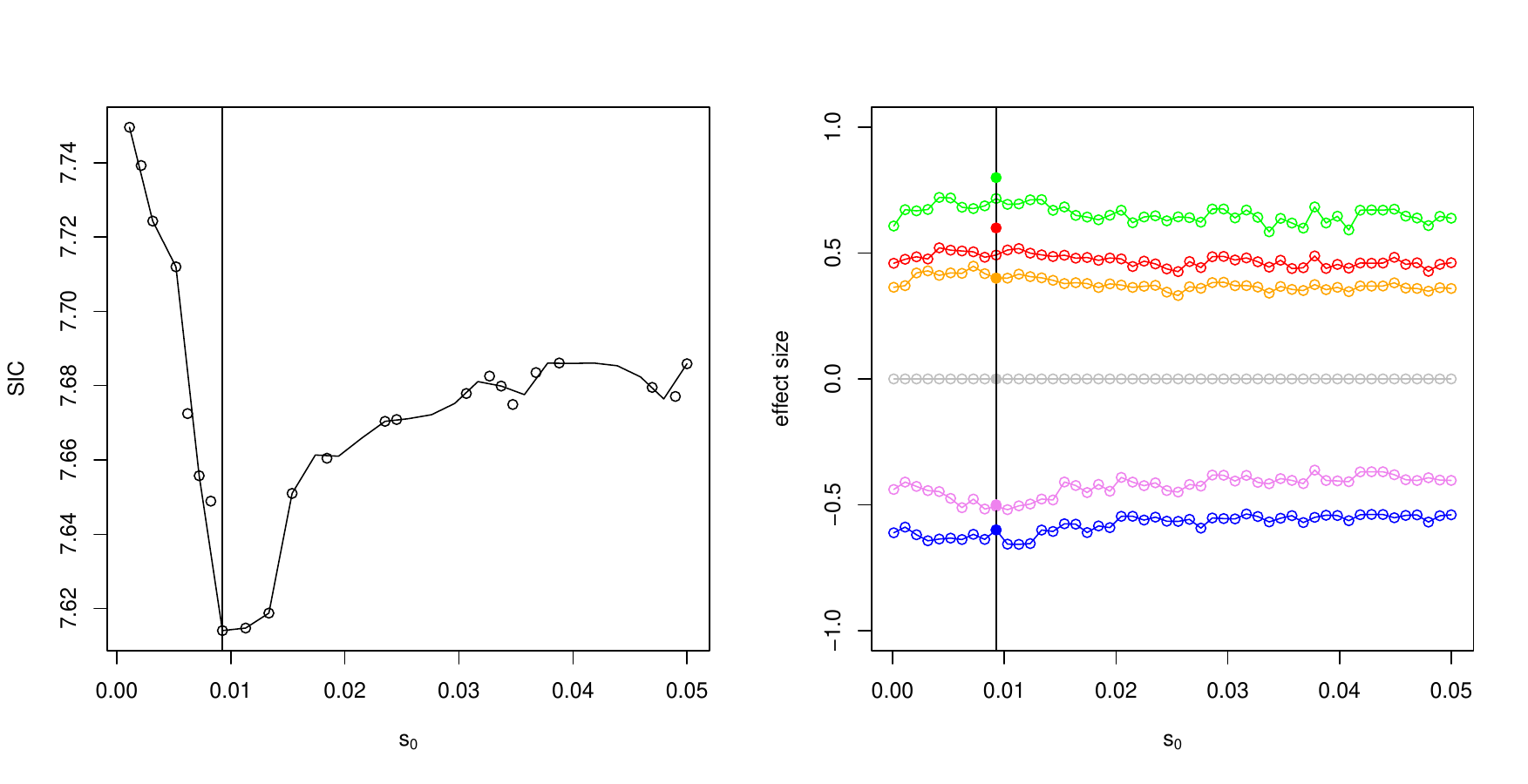}
		\caption{\footnotesize The SIC profile (left) and solution path (right) of ssQLASSO under $\tau=50\%$, AR-1 correlation, $(n=400,p=1600)$ and Error3. The colored circles on the solution path denote the estimated values with respect to the 5 non-zero coefficients represented by filled circles. The vertical lines correspond to the optimal tuning parameter.}\label{BQLSSSP}
	\end{center}
\end{figure}

Besides, for ssQLASSO and LASSO, we have compared their solution paths in Figure \ref{LASSOSP} and Figure \ref{BQLSSSP}, respectively. The response variable has been generated under the setting of ($n,p$)=(400,1600), AR1 correlation and homogeneous Error 3. For better illustration, only 5 nonzero components, $(0.8, 0.6, 0.4, -0.5, -0.6)$, in the coefficient vector $\boldsymbol \beta$ are assumed. We refer you to the subsection of \textbf{Solution path and self-adaptivity} in the Section of Simulation for more details in interpreting Figure \ref{LASSOSP} and Figure \ref{BQLSSSP} to understand the selective shrinkage and self-adaptivity properties of ssQLASSO.



\section{Computational Time}

\begin{table} [H]
	\def\arraystretch{1.5}
	\begin{center}
		\caption{Computational time (in seconds) of homogeneous model under AR-1 correlation, $(n,p)$=(400,1600) with 100 replicates.}\label{rt}
		\centering
		\fontsize{10}{10}\selectfont{
			\begin{tabular}{ l l l l l }
				\hline
				Run time && \multicolumn{1}{c}{$\tau=0.3$}& \multicolumn{1}{c}{$\tau=0.5$} & \multicolumn{1}{c}{$\tau=0.7$} \\
				\hline
				
				Error1 & ssQLASSO & 24.27(4.33) & 28.85(7.65) & 25.27(6.69) \\ 
				& QLASSO & 281.23(10.61) & 183.50(7.73) & 286.97(11.01) \\ 
				& ssLASSO & 9.12(0.43) & 11.66(0.64) & 9.19(0.58) \\ 
				& LASSO & 0.01(0.00) & 0.02(0.00) & 0.01(0.00) \\
				\cline{2-5}
				Error2 & ssQLASSO & 23.08(3.28) & 22.51(3.39) & 24.47(3.97) \\ 
				& QLASSO & 248.85(9.24) & 261.97(13.67) & 275.52(11.42) \\ 
				& ssLASSO & 10.34(0.56) & 9.70(0.68) & 10.80(0.55) \\ 
				& LASSO & 0.02(0.01) & 0.02(0.02) & 0.01(0.00) \\
				\cline{2-5}
				Error3 & ssQLASSO & 27.87(5.30) & 26.14(4.47) & 23.08(3.40) \\ 
				& QLASSO & 293.26(10.98) & 286.14(13.28) & 268.87(10.66) \\ 
				& ssLASSO & 10.82(0.59) & 10.14(0.53) & 10.31(0.48) \\ 
				& LASSO & 0.02(0.02) & 0.02(0.00) & 0.01(0.00) \\
				\cline{2-5} 
				Error4 & ssQLASSO & 24.19(3.84) & 23.53(3.38) & 23.90(4.40) \\ 
				& QLASSO & 264.00(10.66) & 295.07(13.58) & 273.10(10.12) \\ 
				& ssLASSO & 10.43(0.56) & 10.44(0.47) & 10.37(0.25) \\ 
				& LASSO & 0.01(0.00) & 0.02(0.00) & 0.01(0.00) \\
				\cline{2-5}
				Error5 & ssQLASSO & 23.27(3.52) & 24.07(4.34) & 22.94(3.89) \\ 
				& QLASSO & 267.89(10.37) & 292.75(12.99) & 253.00(9.73) \\ 
				& ssLASSO & 10.30(0.35) & 10.36(0.53) & 9.66(0.41) \\ 
				& LASSO & 0.01(0.00) & 0.01(0.00) & 0.02(0.00) \\
				\hline
			\end{tabular}
		}
	\end{center}
	\centering
\end{table}
\newpage 
\section{More Numerical Results}

\subsection{Simulation under large scale settings}

We have conducted additional simulation studies to demonstrate that the ssQLASSO can consistently outperform alternatives with much larger sample size and dimensionality in addition to the cases of $(n,p)=(400,800) \text{and} (400,1600)$ examined before. 
Table \ref{homo2400} and \ref{heter2400} have shown the identification and estimation results when predictors are generated with AR-1 correlation and $(n,p)=(800,2400)$ from the following homogeneous and heterogeneous data generating models:
\begin{equation*}
\begin{split}
\text{homogeneous}: &y_i=2+\boldsymbol x_i^\top\bm{\beta}+\epsilon_i,\\
\text{heterogeneous}: &y_i=2+\boldsymbol x_i^\top\bm{\beta}+(1+x_{i2})\epsilon_i,
\end{split}
\end{equation*}
where details of the data generating mechanism have been described in the Section of Simulation. In addition, using the same data generating model under $(n,p)=(800,3200)$, we have generated the corresponding assessment results in Table \ref{homo3200} and \ref{heter3200} for homogeneous and heterogeneous models, respectively.

In all the aforementioned simulation, we have not included the clinical covariates that are not subject to selection in the data generating model since typically inclusion of low dimensional covariates not subject to selection will not affect the estimation and identification performance on high-dimensional omics predictors. To assess the model performance when clinical covariates are incorporated, we then consider the following data generating models with predictors being simulated with AR-1 correlation and $(n,p)=(800,3200)$ as follows. 
\begin{equation*}
\begin{split}
\text{homogeneous}: &y_i=\boldsymbol z_i^\top\bm{\alpha}+\boldsymbol x_i^\top\bm{\beta}+\epsilon_i,\\
\text{heterogeneous}: &y_i=\boldsymbol z_i^\top\bm{\alpha}+\boldsymbol x_i^\top\bm{\beta}+(1+x_{i2})\epsilon_i,
\end{split}
\end{equation*}
where $\boldsymbol z_{i}=(1,z_{i1},z_{i2},z_{i3})^{\top}$ corresponds to the intercept and 3 clinical covariates, and $\bm{\alpha}=(\alpha_{0},\alpha_{1},\alpha_{2},\alpha_{3})^\top$ with $\alpha_{0}$=2 . The 3 clinical covariates are simulated from multivariate normal distribution and the associated coefficients are generated from the uniform distribution $U$[0.6,0.8]. Table \ref{homoclinic} and Table \ref{heterclinc} show the results under homogeneous and heterogeneous errors, respectively. 
The patterns of model performance in terms of identification and estimations under these additional large scale settings in Tables \ref{homo2400} -- \ref{heterclinc} are consistent with our previous findings, indicating the superiority of ssQLASSO over alternative methods. 

\newpage
\begin{table} [H]
	\def\arraystretch{1.5}
	\captionsetup{font=scriptsize}
\begin{center}
    \caption{Evaluation of homogeneous model under AR-1 correlation, $(n,p)$=(800,2400) with 100 replicates.}\label{homo2400}
    \centering
		\fontsize{7}{7}\selectfont{
\begin{tabular}{ l l l l l l l l }
 \hline
 $n=800$&$p=2400$&& \multicolumn{1}{c}{TP}& \multicolumn{1}{c}{FP} & \multicolumn{1}{c}{F1} & \multicolumn{1}{c}{MCC} & \multicolumn{1}{c}{Estimation}\\
 \hline

$\tau=0.3$& Error1 & ssQLASSO & 15.00(0.00) & 4.33(0.58) & 0.87(0.01) & 0.88(0.01) & 0.67(0.06) \\ 
   &  & QLASSO & 15.00(0.00) & 2.47(1.63) & 0.93(0.05) & 0.93(0.04) & 2.83(0.30) \\ 
   &  & ssLASSO & 15.00(0.00) & 0.57(0.77) & 0.98(0.02) & 0.98(0.02) & 0.43(0.08) \\ 
   &  & LASSO & 15.00(0.00) & 83.43(35.24) & 0.29(0.07) & 0.40(0.06) & 2.45(0.67) \\
\cline{2-8}
& Error2 & ssQLASSO & 15.00(0.00) & 1.13(1.78) & 0.97(0.05) & 0.97(0.05) & 0.71(0.14) \\ 
   &  & QLASSO & 15.00(0.00) & 5.07(2.52) & 0.86(0.06) & 0.87(0.05) & 4.11(0.53) \\ 
   &  & ssLASSO & 13.77(2.21) & 0.00(0.00) & 0.95(0.09) & 0.95(0.08) & 2.03(1.71) \\
\hline
\end{tabular}
}
\end{center}
\centering
\end{table}
\begin{table} [H]
	\def\arraystretch{1.5}
	\captionsetup{font=scriptsize}
\begin{center}
    \caption{Evaluation of heterogeneous model under AR-1 correlation, $(n,p)$=(800,2400) with 100 replicates.}\label{heter2400}
    \centering
		\fontsize{7}{7}\selectfont{
\begin{tabular}{ l l l l l l l l }
 \hline
 $n=800$&$p=2400$&& \multicolumn{1}{c}{TP}& \multicolumn{1}{c}{FP} & \multicolumn{1}{c}{F1} & \multicolumn{1}{c}{MCC} & \multicolumn{1}{c}{Estimation}\\
 \hline

$\tau=0.3$& Error1 & ssQLASSO & 15.00(0.00) & 3.57(2.53) & 0.90(0.07) & 0.90(0.06) & 0.95(0.14) \\ 
   &  & QLASSO & 15.00(0.00) & 2.70(1.49) & 0.92(0.04) & 0.92(0.04) & 2.81(0.36) \\ 
   &  & ssLASSO & 15.00(0.00) & 0.13(0.35) & 1.00(0.01) & 1.00(0.01) & 1.11(0.14) \\ 
   &  & LASSO & 15.00(0.00) & 67.53(25.34) & 0.33(0.08) & 0.43(0.07) & 3.36(0.59) \\
\cline{2-8}
& Error2 & ssQLASSO & 15.00(0.00) & 0.93(2.12) & 0.97(0.06) & 0.98(0.05) & 0.85(0.18) \\ 
   &  & QLASSO & 15.00(0.00) & 4.97(1.90) & 0.86(0.05) & 0.87(0.04) & 3.42(0.54) \\ 
   &  & ssLASSO & 13.47(2.11) & 0.00(0.00) & 0.94(0.09) & 0.94(0.08) & 2.89(1.41) \\
\hline
\end{tabular}
}
\end{center}
\centering
\end{table}

\begin{table} [H]
	\def\arraystretch{1.5}
	\captionsetup{font=scriptsize}
\begin{center}
    \caption{Evaluation of homogeneous model under AR-1 correlation, $(n,p)$=(800,3200) with 100 replicates.}\label{homo3200}
    \centering
		\fontsize{7}{7}\selectfont{
\begin{tabular}{ l l l l l l l l }
 \hline
 $n=800$&$p=3200$&& \multicolumn{1}{c}{TP}& \multicolumn{1}{c}{FP} & \multicolumn{1}{c}{F1} & \multicolumn{1}{c}{MCC} & \multicolumn{1}{c}{Estimation}\\
 \hline

$\tau=0.3$& Error1 & ssQLASSO & 15.00(0.00) & 1.70(1.70) & 0.95(0.05) & 0.95(0.04) & 0.78(0.19) \\ 
   &  & QLASSO & 15.00(0.00) & 3.27(1.98) & 0.90(0.05) & 0.91(0.05) & 2.92(0.27) \\ 
   &  & ssLASSO & 15.00(0.00) & 0.40(0.72) & 0.99(0.02) & 0.99(0.02) & 0.44(0.07) \\ 
   &  & LASSO & 15.00(0.00) & 81.90(28.37) & 0.28(0.07) & 0.40(0.06) & 2.45(0.49) \\
\cline{2-8}
& Error2 & ssQLASSO & 15.00(0.00) & 0.50(0.97) & 0.98(0.03) & 0.98(0.03) & 0.71(0.16) \\ 
   &  & QLASSO & 15.00(0.00) & 6.47(2.05) & 0.83(0.05) & 0.84(0.04) & 3.92(0.48) \\ 
   &  & ssLASSO & 13.50(2.03) & 0.00(0.00) & 0.94(0.08) & 0.95(0.08) & 2.25(1.45) \\
\hline
\end{tabular}
}
\end{center}
\centering
\end{table}
\begin{table} [H]
	\def\arraystretch{1.5}
	\captionsetup{font=scriptsize}
\begin{center}
    \caption{Evaluation of heterogeneous model under AR-1 correlation, $(n,p)$=(800,3200) with 100 replicates.}\label{heter3200}
    \centering
		\fontsize{7}{7}\selectfont{
\begin{tabular}{ l l l l l l l l }
 \hline
 $n=800$&$p=3200$&& \multicolumn{1}{c}{TP}& \multicolumn{1}{c}{FP} & \multicolumn{1}{c}{F1} & \multicolumn{1}{c}{MCC} & \multicolumn{1}{c}{Estimation}\\
 \hline

$\tau=0.3$& Error1 & ssQLASSO & 15.00(0.00) & 2.67(2.47) & 0.92(0.07) & 0.93(0.06) & 0.93(0.12) \\ 
   &  & QLASSO & 15.00(0.00) & 3.47(1.78) & 0.90(0.05) & 0.90(0.04) & 2.82(0.32) \\ 
   &  & ssLASSO & 15.00(0.00) & 0.30(0.79) & 0.99(0.02) & 0.99(0.02) & 1.11(0.15) \\ 
   &  & LASSO & 15.00(0.00) & 82.40(30.43) & 0.29(0.08) & 0.40(0.07) & 3.64(0.69) \\
\cline{2-8}
& Error2 & ssQLASSO & 15.00(0.00) & 0.40(1.04) & 0.99(0.03) & 0.99(0.03) & 0.84(0.14) \\ 
   &  & QLASSO & 15.00(0.00) & 6.60(2.50) & 0.82(0.06) & 0.84(0.05) & 3.69(0.44) \\ 
   &  & ssLASSO & 13.37(1.97) & 0.00(0.00) & 0.94(0.08) & 0.94(0.07) & 2.93(1.31) \\
\hline
\end{tabular}
}
\end{center}
\centering
\end{table}

\begin{table} [H]
	\def\arraystretch{1.5}
	\captionsetup{font=scriptsize}
\begin{center}
    \caption{Evaluation of homogeneous model with clinical factors under AR-1 correlation, $(n,p)$=(800,3200), 100 replicates.}\label{homoclinic}
    \centering
		\fontsize{7}{7}\selectfont{
\begin{tabular}{ l l l l l l l l }
 \hline
 $n=800$&$p=3200$&& \multicolumn{1}{c}{TP}& \multicolumn{1}{c}{FP} & \multicolumn{1}{c}{F1} & \multicolumn{1}{c}{MCC} & \multicolumn{1}{c}{Estimation}\\
 \hline

$\tau=0.3$& Error1 & ssQLASSO & 15.00(0.00) & 3.70(1.97) & 0.89(0.05) & 0.90(0.05) & 0.76(0.15) \\ 
   &  & QLASSO & 15.00(0.00) & 3.87(2.01) & 0.89(0.05) & 0.89(0.05) & 2.85(0.24) \\ 
   &  & ssLASSO & 15.00(0.00) & 0.37(0.76) & 0.99(0.02) & 0.99(0.02) & 0.40(0.08) \\ 
   &  & LASSO & 15.00(0.00) & 82.80(22.16) & 0.28(0.05) & 0.39(0.05) & 2.47(0.38) \\
\cline{2-8}
& Error2 & ssQLASSO & 15.00(0.00) & 0.43(0.97) & 0.99(0.03) & 0.99(0.03) & 0.72(0.15) \\ 
   &  & QLASSO & 15.00(0.00) & 6.47(2.76) & 0.83(0.06) & 0.84(0.05) & 4.11(0.54) \\ 
   &  & ssLASSO & 13.63(2.44) & 0.00(0.00) & 0.94(0.11) & 0.95(0.10) & 2.14(1.83) \\
\hline
\end{tabular}
}
\end{center}
\centering
\end{table}
\begin{table} [H]
	\def\arraystretch{1.5}
	\captionsetup{font=scriptsize}
\begin{center}
    \caption{Evaluation of heterogeneous model with clinical factors under AR-1 correlation, $(n,p)$=(800,3200), 100 replicates.}\label{heterclinc}
    \centering
		\fontsize{7}{7}\selectfont{
\begin{tabular}{ l l l l l l l l }
 \hline
 $n=800$&$p=3200$&& \multicolumn{1}{c}{TP}& \multicolumn{1}{c}{FP} & \multicolumn{1}{c}{F1} & \multicolumn{1}{c}{MCC} & \multicolumn{1}{c}{Estimation}\\
 \hline

$\tau=0.3$& Error1 & ssQLASSO & 15.00(0.00) & 3.63(4.80) & 0.91(0.10) & 0.91(0.09) & 0.88(0.21) \\ 
   &  & QLASSO & 15.00(0.00) & 3.33(1.24) & 0.90(0.03) & 0.91(0.03) & 2.67(0.29) \\ 
   &  & ssLASSO & 15.00(0.00) & 0.07(0.25) & 1.00(0.01) & 1.00(0.01) & 1.11(0.16) \\ 
   &  & LASSO & 15.00(0.00) & 79.57(33.40) & 0.30(0.08) & 0.41(0.07) & 3.65(0.79) \\
\cline{2-8}
& Error2 & ssQLASSO & 15.00(0.00) & 0.87(2.18) & 0.98(0.06) & 0.98(0.05) & 0.87(0.14) \\ 
   &  & QLASSO & 15.00(0.00) & 7.33(2.97) & 0.81(0.06) & 0.82(0.06) & 3.60(0.52) \\ 
   &  & ssLASSO & 13.87(1.91) & 0.07(0.25) & 0.95(0.09) & 0.96(0.08) & 2.92(1.50) \\
\hline
\end{tabular}
}
\end{center}
\centering
\end{table}


\subsection{Computational time under different study sizes}

\begin{figure}[H]
	\begin{center}
		\includegraphics[height=0.45\textheight]{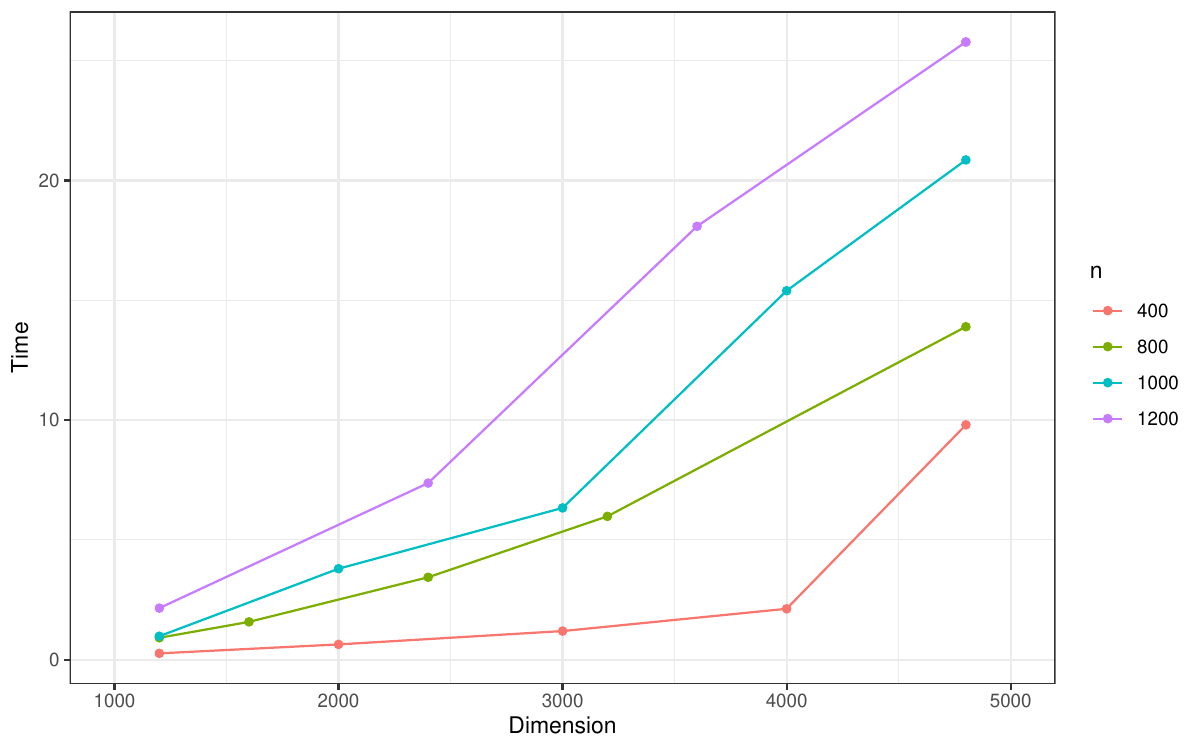}
		\caption{\footnotesize The computational time (in minutes averaged across 100 replicates) of ssQLASSO under $\tau=30\%$, AR-1 correlation and Error2 in terms of $n$ and $p$.}\label{Ctime}
	\end{center}
\end{figure}

The computational times (in minutes) under different combinations of $n$ and $p$ have been demonstrated in Figure \ref{Ctime}.

\newpage
\subsection{Adaptivity of Asymmetric Laplace Distribution}

The proposed ssQLASSO has rooted in quantile regression that have been widely examined in both the frequentist and Bayesian frameworks. In published studies from both frameworks, quantile regression based variable selection methods generally do not require a test to check whether it can be applied since model assessment in high-dimensional settings are mainly based on prediction and identification, as shown in the simulation study and case studies. Nevertheless, it is still necessary to evaluate how well the asymmetric Laplace distribution (ALD) adapts to the widely adopted heavy-tailed errors from published literature. 

As described in Section 2.1, we assume that the model error $\epsilon_{i}$ ($i=1,...,n$) follows the asymmetric Laplace distribution (ALD) with a fixed quantile level $\tau$ and the quantile check loss function \cite{yu2001bayesian}:
\begin{equation*}
f(\epsilon_i|\mu_i,\sigma,\tau)=\frac{\tau(1-\tau)}{\sigma}\text{exp}\left( -\rho_{\tau}\left(\frac{\epsilon_i-\mu}{\sigma} \right)\right),
\end{equation*}
where $\mu$ is the location parameter and $\sigma(>0)$ is the scale parameter to determine the the skewness of ALD. In simulation and case studies, although the distributional assumptions on ALD are violated, the ssQLASSO has still demonstrated superior performance, which is consistent with the observations on robustness of ALD--based likelihood to violations of a diversity of distributional assumptions from published studies \cite{yang2016posterior}. 

\begin{figure}[H]
	\begin{center}
	\hspace*{-4.5cm} 
		\includegraphics[height=0.8\textheight]{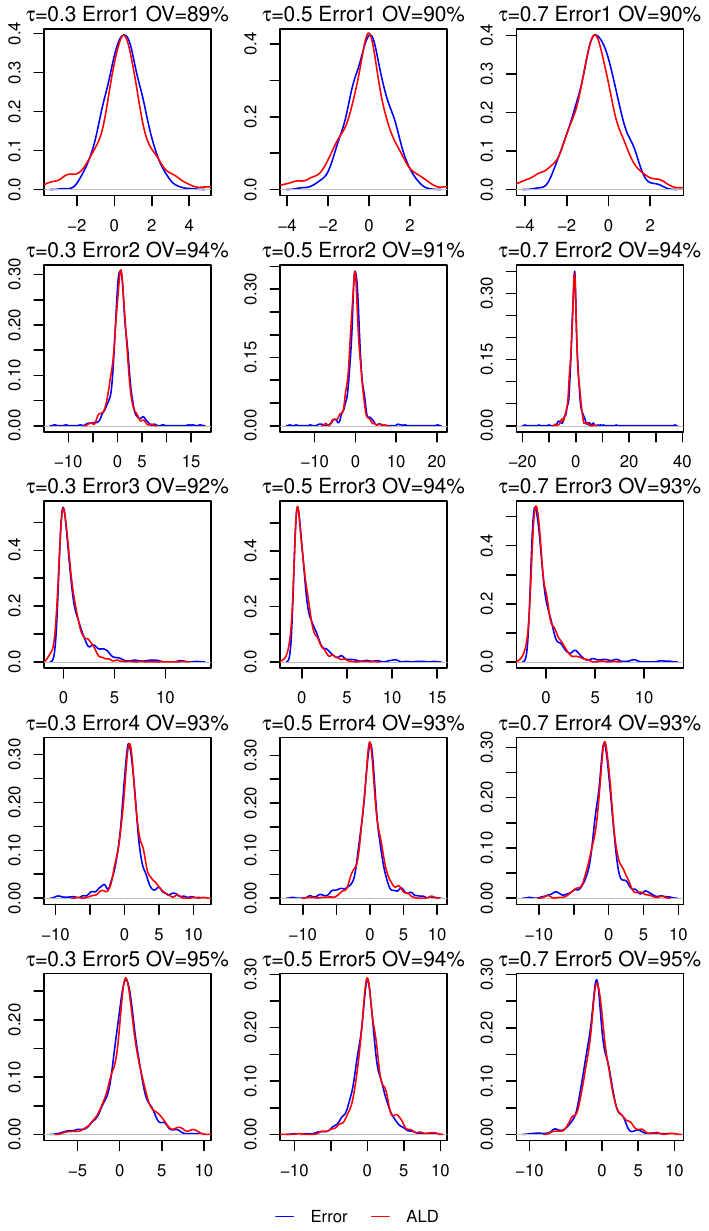}
		\caption{\footnotesize Density plots and overlapping areas of homogeneous error distributions and the corresponding best fitted ALD based on 800 random samples.}\label{density1}
	\end{center}
\end{figure}

\begin{figure}[H]
	\begin{center}
	\hspace*{-4.5cm} 
		\includegraphics[height=0.8\textheight]{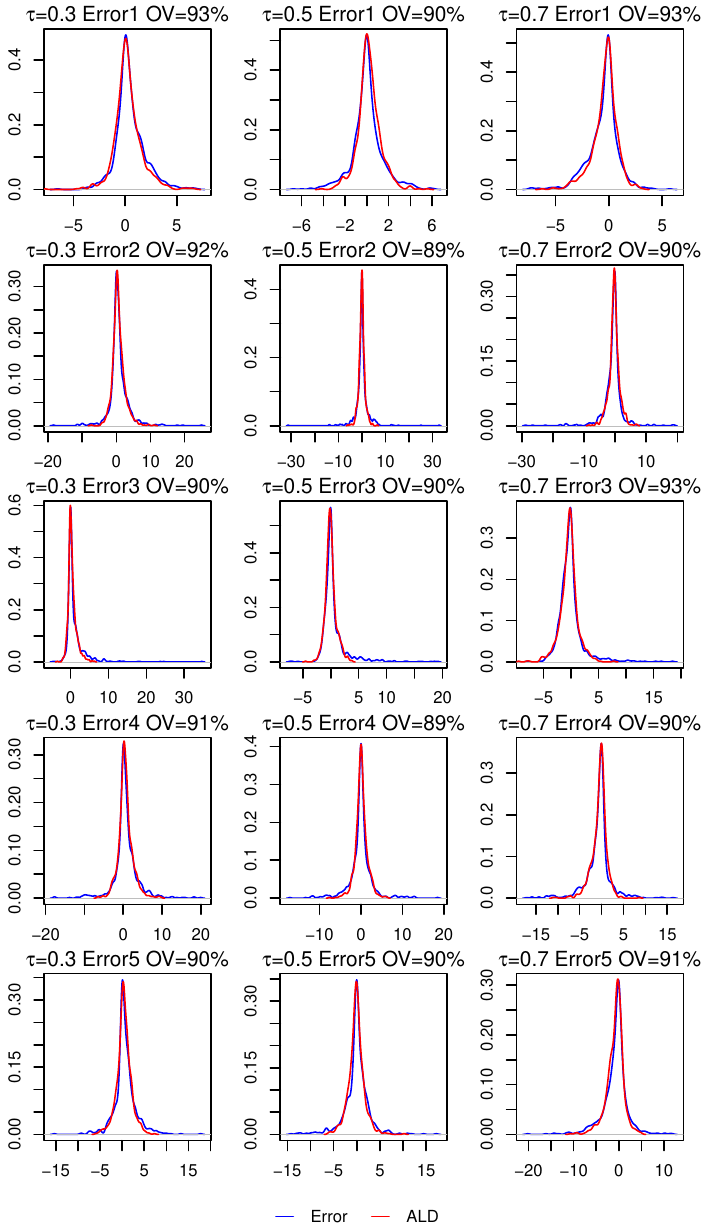}
		\caption{\footnotesize Density plots and overlapping areas of heterogeneous error distributions and the corresponding best fitted ALD based on 800 random samples.}\label{density2}
	\end{center}
\end{figure}

To further justify that the asymmetric Laplace distribution (ALD) is appropriate to modeling a variety of heavy-tailed errors, we have performed a heuristic analysis by checking the maximum overlapping density between the heavy-tailed errors and best fitted ALD which is determined by finding the largest overlapping areas (OV). Figure \ref{density1} and Figure \ref{density2} show that regardless of the type of distribution and quantile level, the overlapping density between ALD and other errors are around or above 90$\%$, indicating that ALD can satisfactorily approximate the target heavy-tailed errors. Therefore, if the location and scale parameters can be accurately estimated, ssQLASSO is expected to yield superior performance as shown in the simulation and case studies. Please note that, Figure \ref{density2} is corresponding to the heterogeneous errors which are highly skewed and not $i.i.d$ even under Error 1 (the normal error). We have provided the R codes to reproduce the analysis.

\newpage

\newpage
\lstset{columns=fullflexible}

\begin{lstlisting}[language=R]
library(ald)
library(rmutil)
library(overlapping)
library(lattice)

set.seed(1)
n = 800
quant = 0.7

# Laplace distribution & homogeneous model
x0 = rlaplace(n, 0, sqrt(2))
x1 = x0 - quantile(x0,probs = quant)

l1 <- seq(0.1,0.9,length.out=20)
l2 <- seq(0.1,0.9,length.out=20)

# Select proper sigma and p at max overlapping area
M <- matrix(0,20,20)
for(i in 1:20){
  for(j in 1:20){
    x2 <- rALD(n,mu=mean(x1),sigma=l1[i],p=l2[j])
    x <- list(x1,x2)
    M[i,j] <- as.numeric(overlap(x))
  }
}
indices <- which(M == max(M), arr.ind=TRUE)
lambda1 <- l1[indices[1]]
lambda2 <- l2[indices[2]]

# Plot density
x2 <- rALD(n,mu=mean(x1),sigma=lambda1,p=lambda2)
plot( density(x1), col="blue",main=paste("OV=",ceiling(100*M[indices]),"%"),xlab="")
lines( density(x2), col="red")
\end{lstlisting}


\newpage
\subsection{Summary statistics of outliers in simulation}

The model errors generated from the simulation have been widely adopted in published literature on high-dimensional robust methods including those based on quantile regressions, such as  \cite{wu2009variable,wang2012quantile,li2010bayesian,ren2023robust,yang2016posterior} among many others. Therefore, it is critical to assess the performance of ssQLASSO on these representative settings which have been used as a common testbed for high-dimensional robust methods. Here, Table \ref{outliers} provides the summary statistics on outliers from data generated from the simulation, where the outliers have been determined by using the built-in R function \emph{boxplot.stats}. We can observe that heterogeneous settings have produced a much larger number of outliers compared to their counterpart settings under the homogeneous errors, including the normal distributions.

\begin{table} [H]
	\def\arraystretch{1.5}
	\begin{center}
		\caption{Number of outliers under 5 error distributions in homogeneous and heterogeneous models across 100 replicates.}\label{outliers}
		\centering
		\fontsize{10}{10}\selectfont{
			\begin{tabular}{ l l l l l }
				\hline
				Count && \multicolumn{1}{c}{$\tau=0.3$}& \multicolumn{1}{c}{$\tau=0.5$} & \multicolumn{1}{c}{$\tau=0.7$} \\
				\hline
				homogeneous & Error1 & 2.98(1.89) & 2.92(2.14) & 2.90(1.93) \\ 
				$n$=400 & Error2 & 32.49(4.91) & 32.83(5.45) & 32.84(5.56) \\ 
				& Error3 & 31.01(4.62) & 30.40(4.84) & 30.68(4.82) \\ 
				& Error4 & 44.04(5.40) & 43.06(5.93) & 43.76(5.66) \\
				& Error5 & 25.05(5.70) & 25.24(6.04) & 24.80(5.61) \\ 
				\cline{2-5}
				$n$=800 & Error1 & 5.82(2.87) & 5.54(2.66) & 5.82(3.06) \\ 
				& Error2 & 66.01(7.74) & 64.74(8.41) & 67.40(8.96) \\ 
				& Error3 & 62.57(6.65) & 61.19(7.17) & 61.81(7.09) \\ 
				& Error4 & 87.22(7.06) & 87.03(7.77) & 87.97(9.80) \\
				& Error5 & 49.42(8.42) & 50.97(7.21) & 49.69(8.15) \\ 
				\hline
				heterogeneous & Error1 & 31.41(6.45) & 42.54(7.49) & 31.99(6.19) \\ 
				$n$=400 & Error2 & 48.06(7.30) & 58.89(7.60) & 49.20(7.17) \\ 
				& Error3 & 49.16(6.57) & 50.82(6.12) & 32.29(6.08) \\ 
				& Error4 & 50.46(7.27) & 58.30(6.63) & 50.71(7.94) \\
				& Error5 & 45.27(7.99) & 58.49(8.09) & 45.56(7.54) \\ 
				\cline{2-5}
				$n$=800 & Error1 & 61.86(10.01) & 83.32(11.07) & 63.45(9.38) \\ 
				& Error2 & 96.09(10.77) & 116.64(10.83) & 96.54(10.17) \\ 
				& Error3 & 97.06(9.38) & 100.53(8.88) & 71.90(7.90) \\ 
				& Error4 & 100.12(10.18) & 119.98(10.90) & 102.78(11.72) \\
				& Error5 & 92.71(10.81) & 117.15(12.53) & 92.24(10.21) \\ 
				\hline
			\end{tabular}
		}
	\end{center}
	\centering
\end{table}

\newpage 
\subsection{Additional real data analysis}

We have also applied the proposed ssQLASSO under quantile level $\tau=0.3$ and $\tau=0.7$. The number of overlapping findings have been provided in Table \ref{real1} and \ref{real2} for LUAD and SKCM, respectively. For LUAD, ssQLASSO results in an PMAD of 14.03 (sd1.87), and an PMSE of 334.73 (sd83.50) under $\tau=0.3$. It leads to an PMAD of 13.97 (sd2.77), and an PMSE of 341.34 (sd128.87) under $\tau=0.7$. For SKCM, the PMAD and PMSE of ssQLASSO ($\tau$=0.3) are 0.75 (sd0.09) and 1.10 (sd0.46), respectively. Under $\tau=0.7$, PMAD and PMSE are 0.77 (sd0.09) and 1.19 (sd0.49). The prediction performances are comparable to ssQLASSO under $\tau=0.5$.

Analysis of LUAD shows that the non-robust ssLASSO has identified 12 genes, ALPL, MSANTD4, IKZF4, SBDS, S100A1, RRN3P3, AGPAT3, DIRC2, POLR3G, FNBP1, ACSF2, TOE1, which have also been detected by at least one of the other 3 methods except DIRC2 and TOE1. However, DIRC2, or Disrupted in renal carcinoma 2, has been found to be associated with the development of renal cancer (Bodmer et al., \cite{bodmer2002disruption} and Savalas et al.,  \cite{savalas2011disrupted}). And TOE1 has been well acknowledged to play a key role in the progression of Pontocerebella Hypoplasia Type 7 (PCH7) (Deng et al., \cite{deng2019toe1} and Lardelli et al. \cite{lardelli2017biallelic}). None of the two genes is related to lung functions. On the other hand, among genes only detected by ssQLASSO method, LUAD related gene SRP14 and SMC3 have been identified by ssQLASSO across all the 3 quantile levels. 

For SKCM, genes MAP4K4, GJA3, INPP1, and KCTD7 have been uniquely found by the proposed ssQLASSO. Among them, the first three (MAP4K4, GJA3 and INPP1) have been detected at $\tau=0.3$, and the last two (INPP1 and KCTD7) have been identified at $\tau=0.7$. As we have analyze in Section of Real Data Analysis, these genes have been verified in scientific literature regarding their roles in the formation and progression of melanoma. The non-robust ssLASSO results in the discovery of 8 genes (ADA, FLJ20373, PER1, LOC126807, COQ9, ARMCX2, ZFAT, KCTD21), including the uniquely identified PER1 and ARMCX2. The plausible associations between the two genes with melanoma have been rarely reported. For example, ARMCX2 has been found as a prognostic marker for gastric cancer \cite{wang2020armcx} and a key player in the acquired methylation in ovarian tumors \cite{zeller2012candidate}.

\begin{table} [H]
	\def\arraystretch{1.5}
	\begin{center}
		\fontsize{9}{9}\selectfont{
			\caption{Identification results for real data analysis. The numbers of genes identified by different methods and their overlaps.}\label{real1}
			\centering
			\begin{tabular}{ l l l l l l l }
				LUAD&&&&&&
				\\
				\hline
				&ssQLASSO(0.3)&ssQLASSO(0.5)&ssQLASSO(0.7)&QLASSO&ssLASSO&LASSO
				\\
				ssQLASSO(0.3)&15 &5 &1 &4 &6 &7
				\\
				ssQLASSO(0.5)& &17 &4 &12 &3 &7 
				\\
				ssQLASSO(0.7)& & &16 &6 &1 &3
				\\
				QLASSO&&&&30 &6 &9  
				\\
				ssLASSO&&&&&12 &9  
				\\
				LASSO&&&&&&51  
				\\
				\hline
			\end{tabular}
		}
	\end{center}
	\centering
\end{table}

\begin{table} [H]
	\def\arraystretch{1.5}
	\begin{center}
		\fontsize{9}{9}\selectfont{
			\caption{Identification results for real data analysis. The numbers of genes identified by different methods and their overlaps.}\label{real2}
			\centering
			\begin{tabular}{ l l l l l l l }
				SKCM&&&&&&
				\\
				\hline
				&ssQLASSO(0.3)&ssQLASSO(0.5)&ssQLASSO(0.7)&QLASSO&ssLASSO&LASSO
				\\
				ssQLASSO(0.3)&10 &4 &3 &1 &2 &4
				\\
				ssQLASSO(0.5)& &13&5 &2 &4 &8 
				\\
				ssQLASSO(0.7)& & &13 &1 &3 &4
				\\
				QLASSO&&&&17 &1 &6  
				\\
				ssLASSO&&&&&8 &4  
				\\
				LASSO&&&&&&27  
				\\
				\hline
			\end{tabular}
		}
	\end{center}
	\centering
\end{table}

\end{document}